\documentclass[10pt,journal,compsoc]{IEEEtran}

\usepackage[english]{babel}
\usepackage[T1]{fontenc}
\usepackage{amsmath}
\usepackage{bm}
\usepackage{hyperref}
\usepackage{graphicx}
\usepackage{bbold}
\usepackage{lineno}
\usepackage{amsfonts}
\usepackage{multirow}
\usepackage[normalem]{ulem}
\usepackage{lipsum}
\usepackage{verbatim}
\usepackage{array}
\usepackage{subfigure}
\usepackage{booktabs}
\usepackage{diagbox}
\usepackage{float} 
\usepackage{makecell}
\usepackage{phaistos}
\usepackage{mathtools}
\usepackage[table,xcdraw]{xcolor}
\usepackage[numbers]{natbib}

\usepackage{tikz}
\usetikzlibrary{quantikz}

\DeclareMathOperator{\tr}{tr}

\DeclareMathOperator{\QTQGQD}{\mathsf{QT-QGQD}}
\DeclareMathOperator{\QTCGQD}{\mathsf{QT-CGQD}}
\DeclareMathOperator{\QTCGCD}{\mathsf{QT-CGCD}}
\DeclareMathOperator{\CTQGCD}{\mathsf{CT-QGCD}}
\DeclareMathOperator{\CTQGQD}{\mathsf{CT-QGQD}}
\DeclareMathOperator{\QTQGCD}{\mathsf{QT-QGCD}}
\DeclareMathOperator{\CTCGCD}{\mathsf{CT-CGCD}}
\DeclareMathOperator{\CTCGQD}{\mathsf{CT-CGQD}}

\newcommand{\vect}[1]{\boldsymbol{#1}}

\begin{document}
\title{Recent Advances for Quantum Neural Networks in Generative Learning}

\author{Jinkai Tian, Xiaoyu Sun, Yuxuan Du,
Shanshan Zhao, Qing Liu, Kaining Zhang, Wei Yi, Wanrong Huang, Chaoyue Wang, Xingyao Wu, Min-Hsiu Hsieh,~\IEEEmembership{Senior Member,~IEEE}, Tongliang Liu,~\IEEEmembership{Senior Member,~IEEE},
Wenjing Yang, Dacheng Tao,~\IEEEmembership{Fellow,~IEEE} 
        
\IEEEcompsocitemizethanks{
\IEEEcompsocthanksitem Jinkai Tian, Wei Yi, and Wanrong Huang are with Artificial Intelligence Research Center, Defense Innovation Institute, 100071 Beijing, China and Tianjin Artificial Intelligence Innovation Center, 300457 Tianjin, China.
\IEEEcompsocthanksitem  Xiaoyu Sun is with The Blackett Laboratory, Imperial College London, London SW7 2AZ, U.K.   
\IEEEcompsocthanksitem  Yuxuan Du, Shanshan Zhao, Chaoyue Wang, Xingyao Wu, and Dacheng Tao are with JD Explore Academy, 101111 Beijing, China.
\IEEEcompsocthanksitem  Qing Liu is with School of Physical and Mathematical Sciences, Nanyang Technological University, 637371, Singapore.
\IEEEcompsocthanksitem Wenjing Yang is with Institute for Quantum Information \& State Key Laboratory of High Performance Computing, College of Computer Science and Technology, National University of Defense Technology, 410073 Changsha, China.  
\IEEEcompsocthanksitem Kaining Zhang, Tongliang Liu are with The University of Sydney, Darlington, NSW 2008, Australia.
\IEEEcompsocthanksitem Min-hsiu Hsieh is with Hon Hai (Foxconn) Research Institute, Taipei, Taiwai.
\IEEEcompsocthanksitem Jinkai Tian and Xiaoyu Sun contributed equally to this work.
\IEEEcompsocthanksitem This work was done when Jinkai Tian and Xiaoyu Sun are research interns in JD Explore Academy. 
\IEEEcompsocthanksitem Correspondence to Yuxuan Du, Wenjing Yang, Xingyao Wu, and Dacheng Tao.
} 
}

\IEEEtitleabstractindextext{%
\begin{abstract}
Quantum computers are next-generation devices  that hold promise to perform calculations beyond the reach of classical computers. A leading method towards achieving this goal is through quantum machine learning, especially quantum generative learning. Due to the intrinsic probabilistic nature of quantum mechanics, it is reasonable to postulate that quantum generative learning models (QGLMs) may surpass their classical counterparts. As such, QGLMs are receiving growing attention from the quantum physics and computer science communities, where various QGLMs that can be efficiently implemented on near-term quantum machines with potential computational advantages are proposed. In this paper, we review the current progress of QGLMs from the perspective of machine learning. Particularly, we interpret these QGLMs, covering quantum circuit Born machines, quantum generative adversarial networks, quantum Boltzmann machines, and quantum autoencoders, as the quantum extension of classical generative learning models. In this context, we explore their intrinsic relation and their fundamental differences. We further summarize the potential applications of QGLMs in both conventional machine learning tasks and quantum physics. Last, we discuss the challenges and further research directions for QGLMs.
\end{abstract}

}

\maketitle

\IEEEdisplaynontitleabstractindextext

\IEEEpeerreviewmaketitle

\IEEEraisesectionheading{\section{Introduction}\label{Sec:Introduction}}
\IEEEPARstart{D}{eep} generative learning models (GLM) have revolutionized the classical world during the past decade~\cite{goodfellow2016deep}, including but not limited to computer vision \cite{voulodimos2018deep}, natural language processing \cite{otter2020survey}, and drug discovery \cite{vamathevan2019applications}. The unprecedented success of GLMs stems from the power of deep neural networks, which can effectively capture the underlying distribution of training data and then generate new samples from the same distribution.   Celebrated by this property, GLMs have been recently exploited to tackle fundamental problems in quantum physics science. Namely, GLMs are used to address the `curse of dimensionality' encountered in quantum physics \cite{carleo2019machine}. Compared with conventional methods, GLMs generally ensure better performance as well as the improved generalization ability. All of these characteristics contribute to physicists to understand the mechanisms of nature.
 
 \begin{figure}[htbp]
    \centering
    \includegraphics[width=0.48\textwidth]{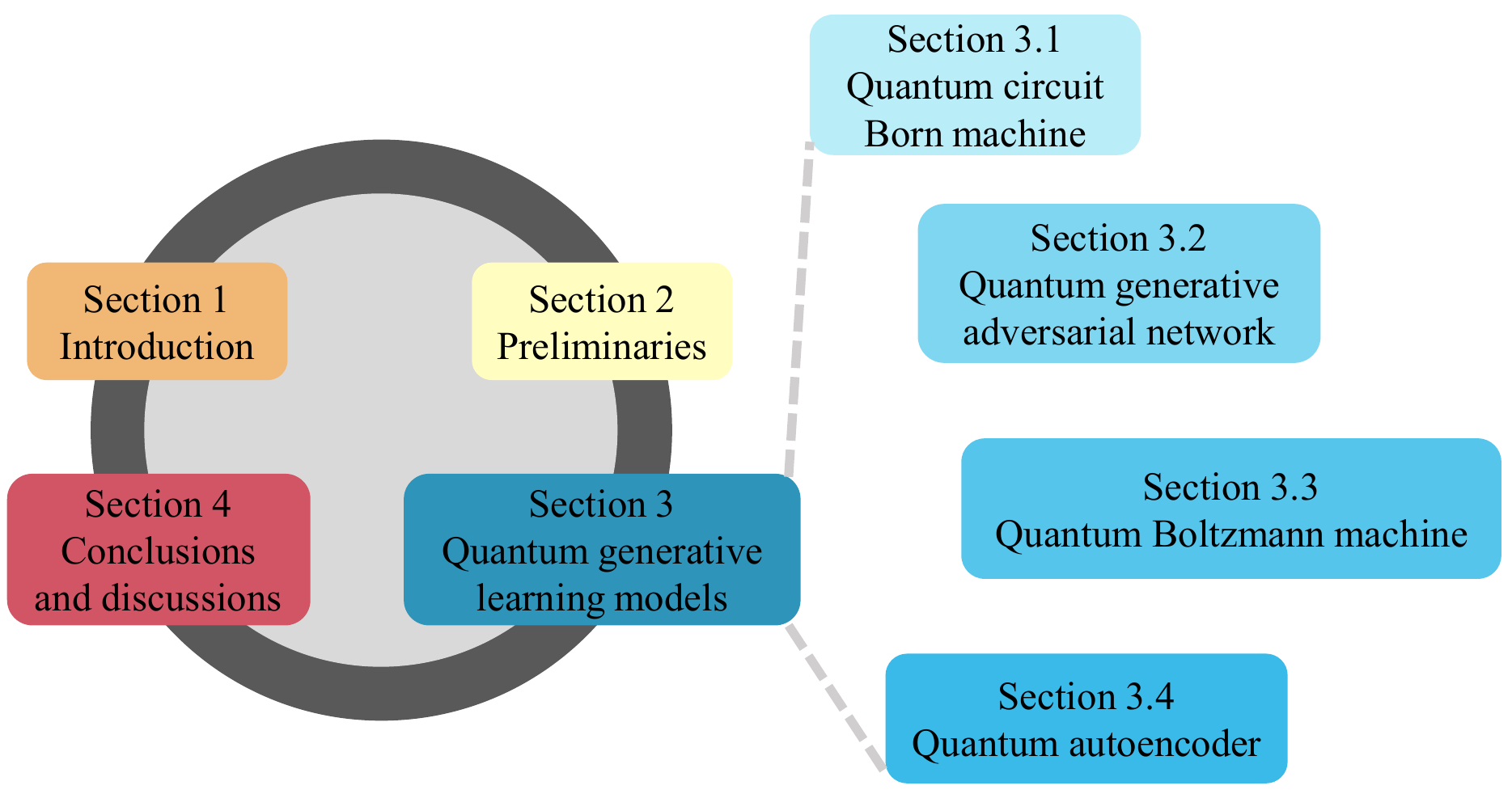}
    \caption{An overview of this survey.}
    \label{fig:sectionpartition}
\end{figure}

In parallel to the design of advanced GLMs and the exploration of their potential applications, another critical line of research in artificial intelligence is seeking the next-generation of GLMs with enhanced abilities. The current challenge is overcoming the computational overhead of GLMs as the limits of Moore’s law is approached \cite{thompson2020computational}. To this end, a leading solution is executing GLMs on quantum computers, which have exhibited strong theoretical and experimental performance \cite{bravyi2020quantum,arute2019quantum}. In this respect, a great amount of effort has been made to design quantum generative learning models (QGLMs) that can be efficiently carried out on noisy intermediate-scale quantum (NISQ) machines \cite{preskill2018quantum} with computational advantages. To date, extensive studies have demonstrated the feasibility of QGLMs for different learning tasks, e.g., image generation~\cite{huang2021experimental}, quantum state approximation \cite{hu2019quantum}, and drug design \cite{jin2022quantum}.

\begin{figure*}[htbp]
    \centering
    \includegraphics[width=0.9\textwidth]{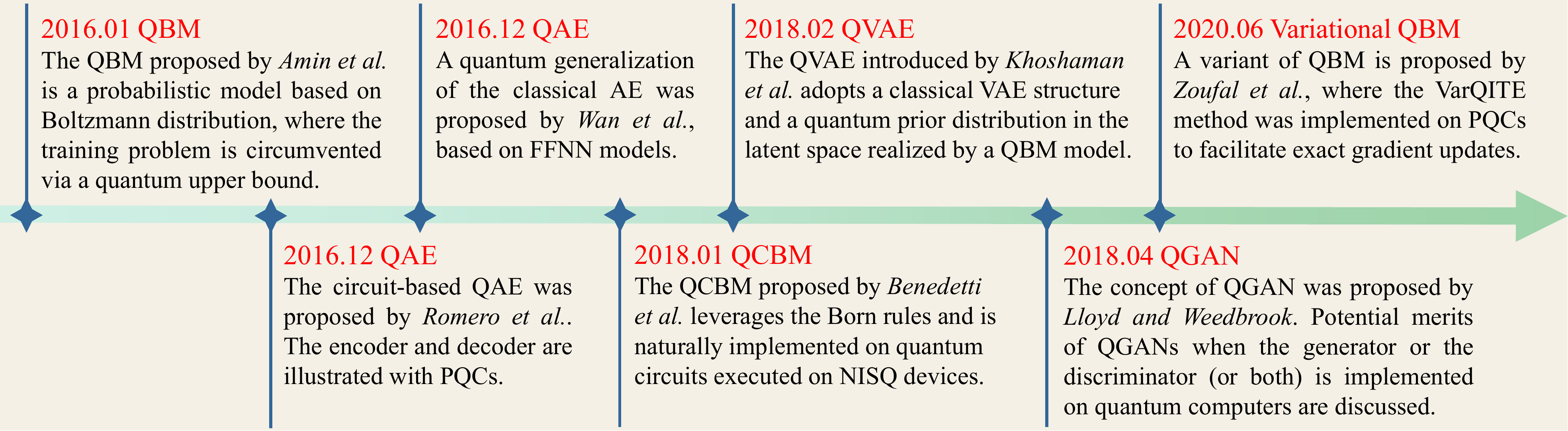}
    \caption{Key landmarks in the development of QGLMs.}
    \label{fig:timeline}
\end{figure*}

The rapid development of  QGLMs necessitates a systematic review of the existing works, which could benefit researchers from both the computer science and quantum physics communities. To this end, in this survey, we analyze the current progress of QGLMs through the lens of deep generative learning. Definitively, according to the typical protocols of GLMs, we categorize QGLMs into four types, which are quantum circuit Born machine (QCBM), quantum generative adversarial network (QGAN), the quantum Boltzmann machine (QBM), and quantum autoencoder (QAE). For each type of QGLM, we first introduce the pioneering work and its inherent relation with its classical counterparts, followed by elucidating its variants and potential applications in both conventional machine learning and quantum physics. To the best of our knowledge, this is the first review in the context of quantum generative learning. We believe this survey can help audiences of varying backgrounds to understand the development of QGLMs.

The structure of this survey is depicted in Fig.~\ref{fig:sectionpartition}. In Section \ref{sec:preliminaries}, we introduce the basic knowledge of deep neural networks, typical classical generative learning models, quantum computing, and variational quantum algorithms. In Section \ref{sec:QGLM}, we systematically review prior literature related to QGLMs and explain the relationship with their classical counterparts. According to the categorization of QGLMs, this section comprises of four subsections, which orient QCBM, QGAN, QBM, and QAE, respectively. In Section \ref{sec:conclusions}, we discuss the challenges and future directions of quantum generative learning.

\subsection{Related works}
A few review articles have similar or intersectional topics with our survey, while their emphases and scopes are different.  \citet{schuld2015introduction},\citet{biamonte2017quantum}, and \citet{ciliberto2018quantum} gave a review on early works of quantum machine learning under fault-tolerant scenarios. Moving into the    NISQ era, \citet{bharti2022noisy}, \citet{cerezo2021variational}, and \citet{massolileap} reviewed the fundamental building blocks of VQAs and their applications. \citet{benedetti2019parameterized} reviewed QNNs as machine learning models for quantum supervised and unsupervised learning tasks.   Although these NISQ-oriented review articles mention quantum generative learning models more or less, none of them provides a comprehensive review of modern QGLMs by elaborating their fundamental mechanism and unveiling the intrinsic relations with GLMs as we did in this survey.

Some reviews working on the quantum extension of discriminative learning models. \citet{li2022recent,ablayev2019quantum,abohashima2020classification} reviewed fault-tolerate quantum classifiers like quantum support vector machines \cite{rebentrost2014quantum} and variational quantum classifiers utilizing QNN in NISQ era. Contrary to these studies concentrating on discriminative learning, our survey focuses on quantum generative learning models. Another branch of surveys is summarizing how classical generative learning models can be used in understanding fundamental problems in quantum physics \cite{carleo2019machine,dunjko2018machine}. There are also several reviews on specific applications, e.g., \citet{herman2022survey} for financial applications, \citet{fedorov2022vqe} for VQE, and \citet{guan2021quantum} for high energy physics. All of these reviews can be regarded as the complement of our survey.

\section{Preliminaries}
\label{sec:preliminaries}
To accommodate the diverse backgrounds of readers, this section presents a brief introduction of neural networks, classical generative learning models, quantum computing, and variational quantum algorithms. Unless explicitly stated, the notations used in this survey are summarized in Table~\ref{table:notations} and the frequently used acronyms are summarized in Table~\ref{table:acronyms}.

\begin{table}[htbp]
	\renewcommand\arraystretch{1.1}
	\caption{Summary of Notations}
	\label{table:notations}
    \begin{tabular}{l p{6cm}}
        \toprule    
        \textbf{Notation} & \textbf{Description} \\    
        \midrule   
        $\vect{x},\vect{y},\vect{z}$ &  Vectors    \\   
        $\vect{A},\vect{B},\vect{C}$ &   Matrices or tensors \\
        $\vect{A}^{\dagger}$  & Hermitian adjoint of a matrix $\vect{A}$   \\ 
        $\mathcal{X},\mathcal{Y},\mathcal{Z}$  & Sets   \\ 
        $\otimes$  & Tensor products, e.g., $|0\rangle^{\otimes n}$ means tensor products of n qubits in state $|0\rangle$ \\ 
        $\tr(\cdot)$  &  Trace operation \\ 
        $\mathbb{1}_n$ & An $n \times n$ identity matrix \\
        $\{\Pi_i\}$  &  Measurement operator \\ 
        $p_{\vect{\theta}}$   & A generated distribution with the tunable parameters $\vect{\theta}$ \\ 
        $q$   & A target distribution   \\ 
        $f \circ g (\vect x)$  & Function composition \\
        $\mathcal{C}_{model}$  &  Cost function of a model   \\  
        $U(\vect\theta)$  & A unitary operator (quantum gate) with parameters  \\ 
        $\sigma_x$, $\sigma_y$, $\sigma_z$ & Standard Pauli matrices   \\ 
        $RX$, $RY$, $RZ$   & Single-qubit rotation operators \\   
        $\mathbb{E}_{\vect{x} \sim P(\vect{x})}[f(\vect{x})]$ &   Expectation values of $f(\vect x)$ for random variable $\vect{x}$ following distribution $P(\vect x)$\\
        \bottomrule   
    \end{tabular}  
\end{table}
 
\begin{table}[htbp]
	\renewcommand\arraystretch{1.1}
	\caption{Summary of Acronyms}
	\label{table:acronyms}
    \begin{tabular}{ll}
        \toprule    
        \textbf{Terminology} & \textbf{Acronym} \\ 
        \midrule      
        Quantum generative learning model  & QGLM \\
        Quantum circuit Born machine & QCBM \\
        Quantum generative adversarial network & QGAN \\
        Quantum Boltzmann machine & QBM \\
        Quantum autoencoder & QAE \\
        Quantum variational autoencoder & QVAE \\
        Quantum neural network & QNN \\
        Variational quantum algorithm & VQA \\
        Parameterized quantum circuit & PQC \\
        Noisy near-term intermediate-scale quantum & NISQ \\
        Quantum approximate optimization algorithm & QAOA \\
        Matrix product state & MPS \\
        Multi-layer perceptron & MLP \\
        Tree tensor network  & TTN \\
        Neural-network quantum state & NQS \\
        Quantum state tomography & QST \\
        Artificial neural network & ANN \\
        Feedfoward neural network  & FFNN \\
        Generative adversarial network & GAN \\  
        Restricted Boltzmann machine & RBM \\
        Variational autoencoder & VAE \\
        Quantum evidence lower bound & Q-ELBO \\
        semi-restricted QBM & semi-RQBM\\ 
        \bottomrule   
    \end{tabular}  
\end{table}

\begin{figure*}[htbp]
    \centering
    \includegraphics[width=1\textwidth]{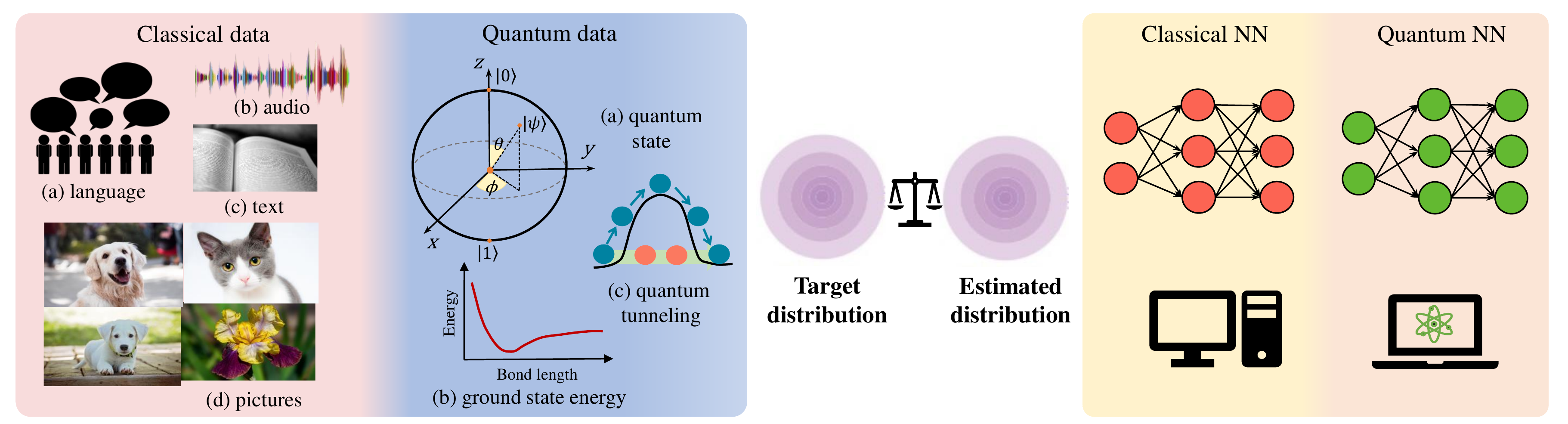}
    \caption{
    An overview of classical and quantum generative learning models.
    The left panel illustrates data distributions of interest in both classical and quantum generative learning.
    The right panel illustrates the similar working mechanism for both classical and quantum generative learning models (i.e., implemented by classical neural networks and quantum neural networks). Concisely, both learning models aim to minimize the discrepancy between the target distribution and the estimated distribution they generate. The process of minimization is completed by a classical optimizer, which continuously updates the trainable parameters of the learning model.} 
    \label{fig:overview} 
\end{figure*}

\subsection{Neural networks}

Although machine learning has experienced challenges and setbacks in the past several decades, it has witnessed a recent resurgence because of the resounding success of neural networks. Neural networks are a type of machine learning framework, which are also called artificial neural networks (ANN). Inspired by the human nervous system, neural networks are designed to implement machine learning algorithms to process various type of data, including images, text, and speech. Relying on the advent of the relevant learning theory and hardware resources, we have witnessed the remarkable development of neural networks during the past decade. Nowadays, neural networks are extensively applied in various fields of artificial intelligence, including computer vision and natural language processing, which enable the research community to explore solutions for a plethora of challenging problems in the real world. In this subsection, we give a brief introduction to neural networks, beginning with the \textit{perceptron}--- a single-layer neural network. 

Conceptually, the perceptron is a linear binary classifier, which has only one neuron. From the mathematical perspective, the perceptron is a linear function and thus can only process linearly separable data patterns. This limited capacity for modeling complex patterns hinders its real-world application. By concatenating a series of perceptrons and non-linear activation functions, the neural network, dubbed multi-layer perceptron (MLP) or feedforward neural network (FFNN), can achieve an enhanced capacity. Since MLP was proposed, it has been widely applied to different learning tasks, including image classification, machine translation, and speech recognition. The key components of an MLP are an input layer, an output layer, and some hidden layers. The input layer does not have learnable parameters, while the output and hidden layer contain a perceptron and a non-linear operation. An intuition of MLP is shown in the right subfigure of Fig.~\ref{fig:overview} (classical NN). The mathematical formulation of MLP yields
\begin{equation}
    \label{eq:nn}
    \vect{y} = f_{\vect{w}} (\vect{x}) = \mathcal{L}^{(L)} \circ \cdots \circ \mathcal{L}^{(1)}(\vect{x}),
\end{equation}
where $\vect{x}$ and $\vect{y}$ are the input and output, respectively, and $\vect{w} = \{ \vect{W}^{(1)}, \vect{b}^{(1)},\dots,\vect{W}^{(L)}, \vect{b}^{(L)}\}$ is the set of trainable parameters. More specifically, $\vect{W}^{(i)} \in \mathbb{R}^{d_i \times d_{i-1}}$ and $\vect{b}^{(i)} \in \mathbb{R}^{d_i}$ are trainable parameters of the $i$-th layer and $d_i$ refers to the dimension of the $i$-th layer. Each layer $\mathcal{L}^{(i)}:\mathbb{R}^{d_{i-1}} \longrightarrow \mathbb{R}^{d_{i}}$ consists of a linear transformation followed by a non-linear activation function $\phi(\cdot)$, i.e.,
\begin{equation}
\vect{x}^{(i)} = \mathcal{L}^{(i)}(\vect{x}^{(i-1)})= \phi (\vect{W}^{(i)} \vect{x}^{(i-1)}+\vect{b}^{(i)}),
\end{equation}
where $ i \in \{1,2,\dots,L\}$ and $\vect{x}^{(0)} = \vect{x}$. For the hidden layer, $\phi(\cdot)$ can be ReLU, sigmoid, or tanh function, while for the output layer $\phi$ can be softmax for classification or sigmoid for regression. The weights are initialized randomly and the MLP can be trained with data through backpropagation.  The cost function which represents the discrepancy between the predicted results and the true label is designed to provide gradient information for parameter updates. The universal approximation theorem \cite{hornik1989multilayer} shows that an MLP with enough parameters can approximate an arbitrary function on a bounded subset of $\mathbb{R}^n$ to any degree of accuracy.

Apart from MLP, there are many other neural networks designed for specific data/tasks. For example, recurrent neural networks (RNN)~\cite{rumelhart1985learning} are exploited to process the sequential data, like text data. Convolutional Neural Networks (CNN) with the weight sharing architecture~\cite{yamashita2018convolutional} are proposed for image data analysis. Recently, the Transformer~\cite{vaswani2017attention} consisting of MLPs has dominated the natural language processing (NLP) area~\cite{devlin-etal-2019-bert,yang2019xlnet} and has further been extensively studied in the computer vision community~\cite{dosovitskiy2020image,liu2021swin}. Celebrated by the remarkable development of Graphics Processing Units (GPUs), we can effectively train a neural network with over one billion parameters on a large amount of data.  For example,  in NLP, GPT~\cite{radford2018improving} contains around 0.1 billion parameters and is trained on 5GB of data. Following the same manner, GPT-2~\cite{radford2019language} is designed with over 1 billion parameters, and GPT-3~\cite{brown2020language} has more parameters, at over 100 billion, and is learned on very large-scale text data, of around 45TB. In computer vision, the model usually contains fewer parameters than the NLP model. For example, a typical large-scale vision model, ConVit~\cite{d2021convit} has 145 million parameters.  

\subsection{Classical generative learning models}
\label{sec:TGLM}

There are primarily three learning paradigms in machine learning : discriminative learning, generative learning, and reinforcement learning (RL). As illustrated in Fig.~\ref{fig:overview}, generative learning models aim to find the underlying distribution of the training dataset and generate new samples of this distribution. For ease of understanding the quantum generative learning models, this subsection separately introduces four representative classical generative learning protocols, i.e., multi-layer perceptron, generative adversarial network, Boltzmann machine, and (variational) autoencoder.

\subsubsection{Multi-layer perceptron generative network}\label{subsec:MLP-GLM}
Though the MLP is widely applied in the field of classification and regression  \cite{murtagh1991multilayer}, it could be interpreted as the simplest generative model for estimating discrete distributions as well. Let us recall the description of MLP in Eqn.~\eqref{eq:nn}, the input data $\vect{x} \in \mathbb{R}^{d_0}$ is mapped to the output $\vect{y}=f_{\bm{w}}(\vect{x}) \in \mathbb{R}^{d_L}$. When MLP is applied to estimate a discrete distribution $q$ with in total $n$ possible outcomes, the input data is keeping fixed during the training and the dimension of the output layer is set as $d_L=n >1 $. The normalized output $\vect{y}$ amounts to the estimated discrete distribution $p_{\bm{w}}$. For instance, when the softmax function is applied to the last layer, the output satisfies $\sum_{i}{y}_i=1$, where for $\forall i \in [n]$, the element ${y}_i$ corresponds to the probability of sampling $i$, i.e., $p_{\vect w}(i)=y_i$.  
\begin{figure}[htbp]
    \centering
    \includegraphics[width=0.4\textwidth]{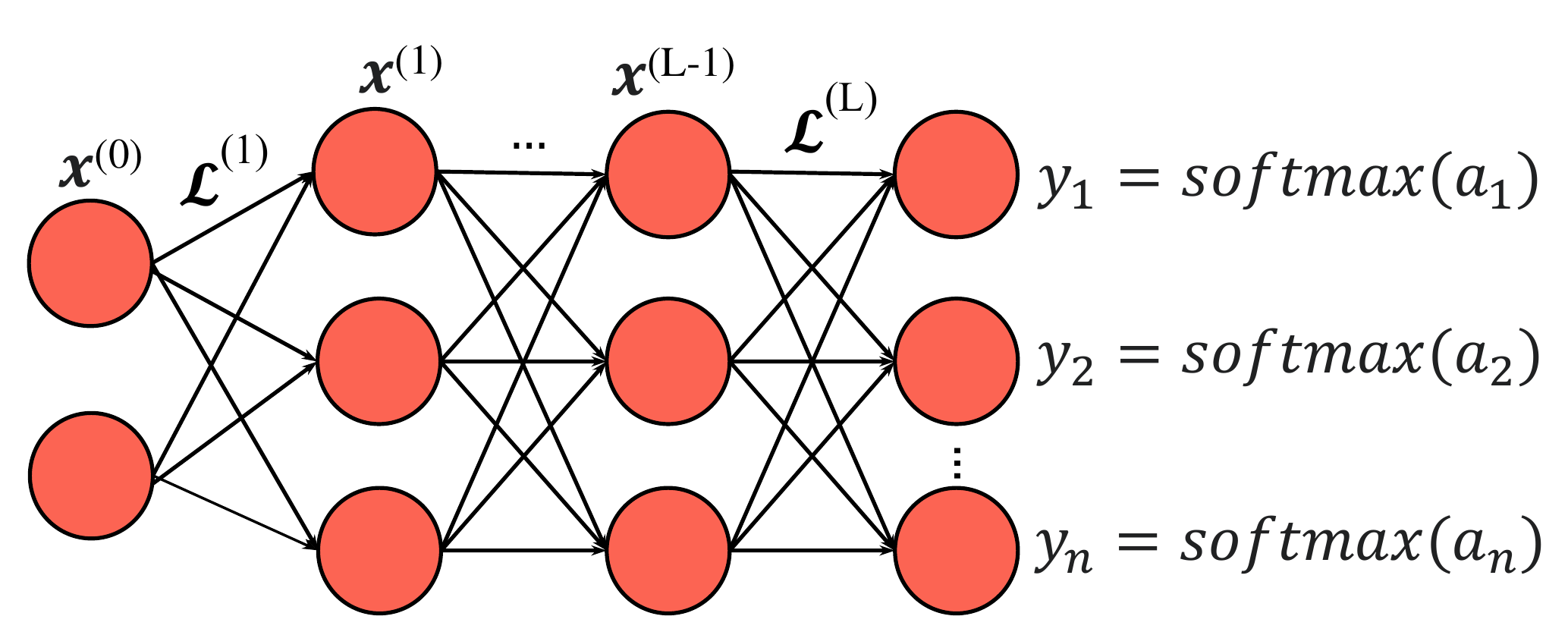}
    \caption{A generative learning model implemented by MLP in the task of discrete distribution estimation. The output layer of MLP corresponds to the estimated distribution. The value of the $i$-th neuron stands for the probability of sampling the outcome $i$. For simplicity, we use the notation $\vect{a}=\vect{W}\vect x^{L-1}+\vect b$, and the softmax function is $softmax(a_i)=e^{a_i}/\sum_{i=1}^n e^{a_i}$.}
    \label{fig:MLP}
\end{figure} 
Let $\mathcal{D} =\{{x}^{m}\}$ be the set of training examples sampled form $q$, which forms the empirical distribution $\hat{q}$. In the training procedure, MLP updates the trainable parameters $\bm{w}$ to minimize the distance between $\hat{q}$ and $p$. This can be completed by minimizing the negative log-likelihood cost function $\mathcal{C}_{MLP}=-\frac{1}{|\mathcal{D}|} \sum_{m}\log[p_{\vect{w}}({x}^{m})]$. When the number of training examples is sufficient large and the cost function approaches to the minimum, $p_{\bm{w}}$ is close to $q$ with a   low estimation error.

\subsubsection{Generative adversarial network}
\label{subsec:classical_GAN}

Since the seminal work of \citet{goodfellow2014generative}, Generative Adversarial Networks (GANs) have been extensively studied due to their powerful ability to simulate complex target distributions.  Different from most generative models adopting the unsupervised learning manner, GAN provides a novel adversarial learning scheme to frame the problem as a minimax training problem by introducing a discriminator, the adversary of the generator. Benefiting from their superiority, GANs have been exploited in a myriad of fields, such as image-to-image translation tasks~\cite{isola2017image,wang2018Perceptual,zhu2017unpaired}, image generation~\cite{bao2017cvae}, and domain adaptation~\cite{tzeng2017adversarial}.

\begin{figure}[htbp]
    \centering
    \includegraphics[width=0.48\textwidth]{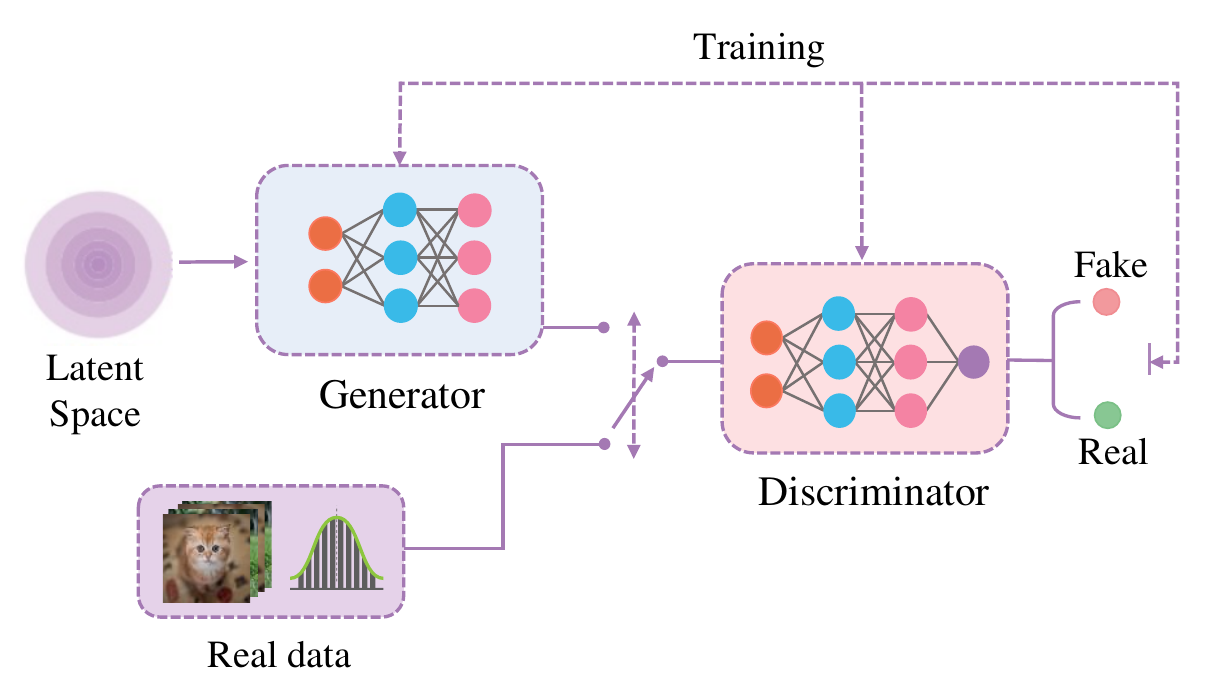}
    \caption{GAN models workflow. The generator generates fake data from some random inpcannotuts to imitate real-world data and tries to deceive the discriminator. The discriminator learns to find the fundamental difference between the generated data and the real data. When the loss converges towards the Nash equilibrium, the generator can generate data that cannot be distinguished by the discriminator.
    }
    \label{fig:GAN}
\end{figure}
 
A schematic workflow of GANs is exhibited in Fig.~\ref{fig:GAN}. There are two main components during the construction and training of a generative adversarial network: a generator $G$ and a discriminator $D$. The generator receives a random input vector $\vect{z} \sim p_z(\vect{z})$ (i.e., $p_z$ refers to a prior distribution like the uniform distribution or the Gaussian distribution) and generates data $\vect{x}=G(\vect{z})$ with learnable parameters $\vect{w_G}$. The goal of the generator is to simulate the target data distribution $q$. The real data sampled from the target distribution and the generated data are fed into the discriminator $D$ with parameters $\vect{w_D}$. The goal of the discriminator is to distinguish data from these two different sources by labeling the target data and the generated data with $1$ and $0$, respectively. In the learning procedure, $\vect{w_G}$ is optimized to generate data in a pattern mimicking the target data distribution, and $\vect{w_D}$ is optimized in the direction of assigning the right label for the target data $\vect{x} \sim q(\vect{x})$ and the generated data $\vect{x} \sim G(\vect{z})$, respectively. Mathematically, the two-player minimax game takes the form $ \min_{\vect{w_G}} \max_{\vect{w_D}} \mathcal{C}_{\text{GAN}} (\vect{w_G}, \vect{w_D})$, where the explicit form of the cost function $\mathcal{C}_{\text{GAN}} (\vect{w_G}, \vect{w_D})$ is 
\begin{equation}
	\label{gan}
	\begin{aligned}
		   \mathbb{E}_{\vect{x} \sim q(\vect{x})}[\log D(\vect{x})] +\mathbb{E}_{\vect{z} \sim p_z(\vect{z})}[\log (1-D(G(\vect{z})))].
	\end{aligned}
\end{equation}
When the training strategy reaches the Nash equilibrium, the discriminator cannot distinguish the generated data from the target data.  

Optimizing such a minimax objective function could be a challenging task since the parameters of both generator and discriminator are dynamically  changed. For this reason, it is easy to encounter mode collapse and gradient vanish issues during GANs' training.  Many variants of the vanilla GAN \cite{goodfellow2014generative} are proposed to tackle specific problems with improved performance. For example, Deep Convolutional GAN (DCGAN) \cite{DBLP:journals/corr/RadfordMC15} implements the generator and the discriminator by using convolutional neural networks and de-convolutional neural networks respectively, which enhances the training stability. By involving extra information in the generator and the discriminator, e.g., class label, the vanilla GAN is transformed into the conditional GAN (CGAN)~\cite{DBLP:journals/corr/MirzaO14}, which has the ability to control the outcome of the generator. To alleviate the vanishing gradient problem and overcome the mode collapse obstacle, Wasserstein GAN (WGAN) \cite{arjovsky2017wasserstein} proposes to employ a cost function derived from the Wasserstein distance. Taking advantage of the coarse-to-fine scheme for high-quality image generation, Laplacian GAN (LapGAN) \cite{10.5555/2969239.2969405} develops a sequential image generation framework by adopting the Laplacian pyramid coding \cite{burt1987laplacian} and conducting up-sampling and down-sampling for the generator and the discriminator respectively. By combining different objectives of GANs with the evolutionary strategy, Evolutionary GAN~\cite{Wang2019Evolutionary} not only alleviates the mode collapse issue, but further improved the generation performance. To explore the properties of unlabelled data, Information Maximizing GAN (InfoGAN) \cite{chen2016infogan} introduces a regularization term, i.e., the mutual information, to the cost function, and improves the interpretability of input vectors for the generator.  By introducing a style-based generator, StyleGAN~\cite{karras2019style} is able to disentangle the feature representations and learn high-level attributes, which benefits from having controll over the image generation process. For large-scale and high-quality image generation, Big Generative Adversarial Network (BigGAN) \cite{brock2018large} improves previous works concerning the training strategy and network structures.  For the recent improvements and trends of classical GANs, refer to Ref.~\cite{gui2021review,jabbar2021survey,pan2019recent}.
\subsubsection{Boltzmann machine}\label{sec:clc_bm}

\begin{figure}[htbp]
    \centering
    \includegraphics[width=0.5\textwidth]{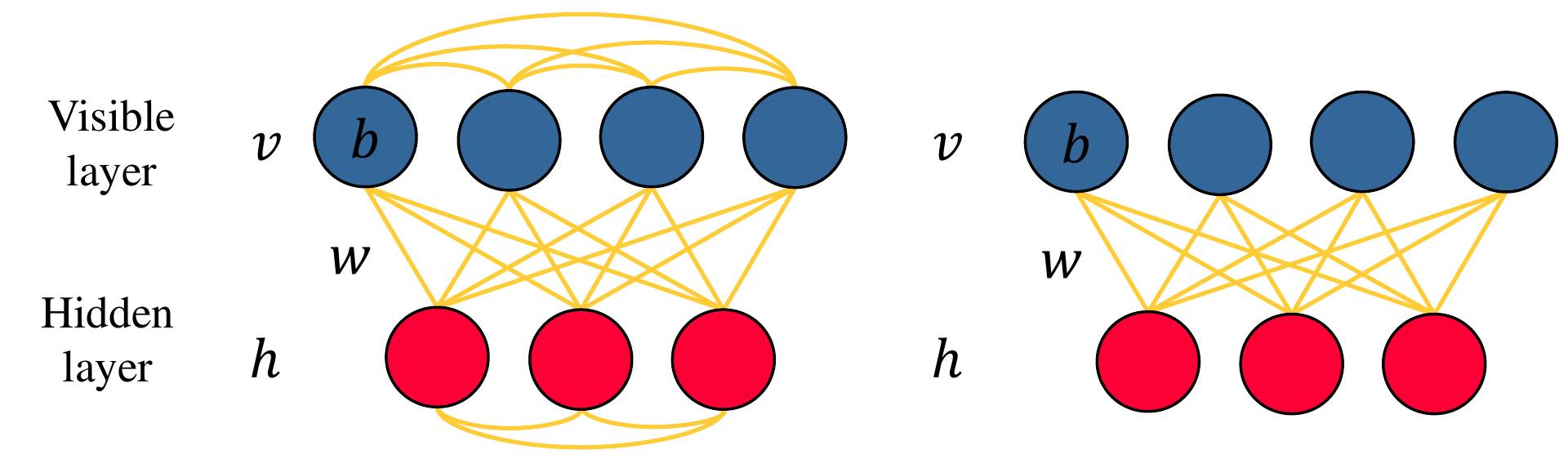}
    \caption{Paradigm of a fully connected BM (left) and a RBM (right). In a fully connected BM, units of visible layer $\vect{v}$  and hidden layer $\vect{h}$ are coupled both internally and externally with interaction weights $w$. In an RBM, internal connections in both the visible and hidden layers are forbidden. The term  $b$ refers to the local bias of each unit.  }
    \label{fig:BM}
\end{figure}

A Boltzmann machine (BM) \cite{ackley1985learning} is essentially a stochastic Ising model with an external field. Mathematically, a Boltzmann machine contains an array of visible units, $\vect{v}$, and an array of hidden units, $\vect{h}$. For simplicity, here, we only discuss the binary-valued BM, which means that all units $z_i \in \vect{z}=\{\vect{v}, \vect{h}\}$ are binary variables (i.e., $z_i =\pm 1$). As a type of energy-based model (EBM)~\cite{lecun2006tutorial}, the cost function of BM corresponds to the energy function, i.e.,
\begin{equation}
\label{eq:ising}
E_{\vect{z}} = -\sum_{i<j} w_{i j} z_{i} z_{j} - \sum_{i} b_{i} z_{i},
\end{equation}
which can also be represented in a matrix product form as $E_{\vect{z}} = - \vect{z}^T \vect{W} \vect{z} - \vect{b}^{T} \vect{z}$ with $\vect{W}$ ($w_{ij}$) and $\vect{b}$ ($b_i$) being trainable parameters. According to the Boltzmann distribution, the probability of observing a visible state $\vect{v}$ can be calculated by summing over the hidden variables, 
\begin{equation}\label{bm-prob}
p_{\vect{\theta}}(\vect{v})=Z^{-1} \sum_{\vect{h}} e^{-E_{\{\vect{v,h}\}}}, \quad Z=\sum_{\vect{z}} e^{-E_{\vect{z}}}.
\end{equation}
Therefore, the parameters of the Boltzmann machine $\vect{w} = \{ \vect{W}, \vect{b}\}$ could be optimized towards the direction in which the generated distribution $p_{\vect{w}}$ is closer to the real distribution of dataset $q$ by performing gradient ascent~\cite{ackley1985learning}.  However, due to the expensive computational overhead, this learning strategy is usually impractical, especially when the number of hidden units is large. To overcome this issue, restricted Boltzmann machines (RBMs),  as shown in Fig.~\ref{fig:BM} are introduced~\cite{smolensky1986information}, where the connections between the hidden and visible units are allowed while the connections within the units, both visible and hidden, are restricted. The RBM can be optimized efficiently via contrastive divergence~\cite{hinton2002training} with Gibbs sampling since the update of hidden and visible units can be implemented in parallel.

\subsubsection{(Variational) Autoencoder}\label{sec:clc_ae}
\begin{figure}[ht]
    \centering
    \subfigure [The layout of a classical AE.]
    {
    \includegraphics[width=0.43\textwidth]{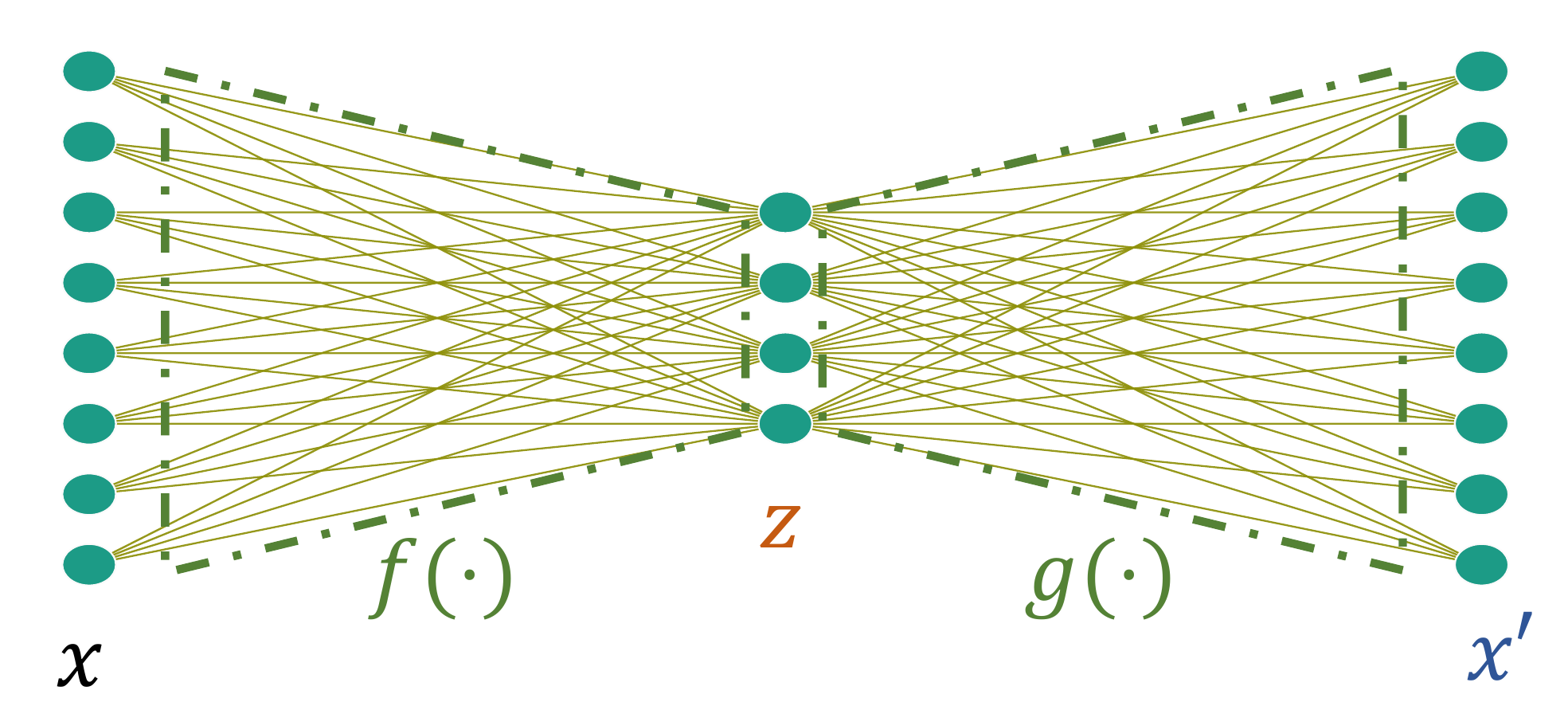}
    \label{fig:AE}
    }
    \vskip\baselineskip

    \subfigure[The layout of a VAE.] 
    {
    \includegraphics[width=0.43\textwidth]{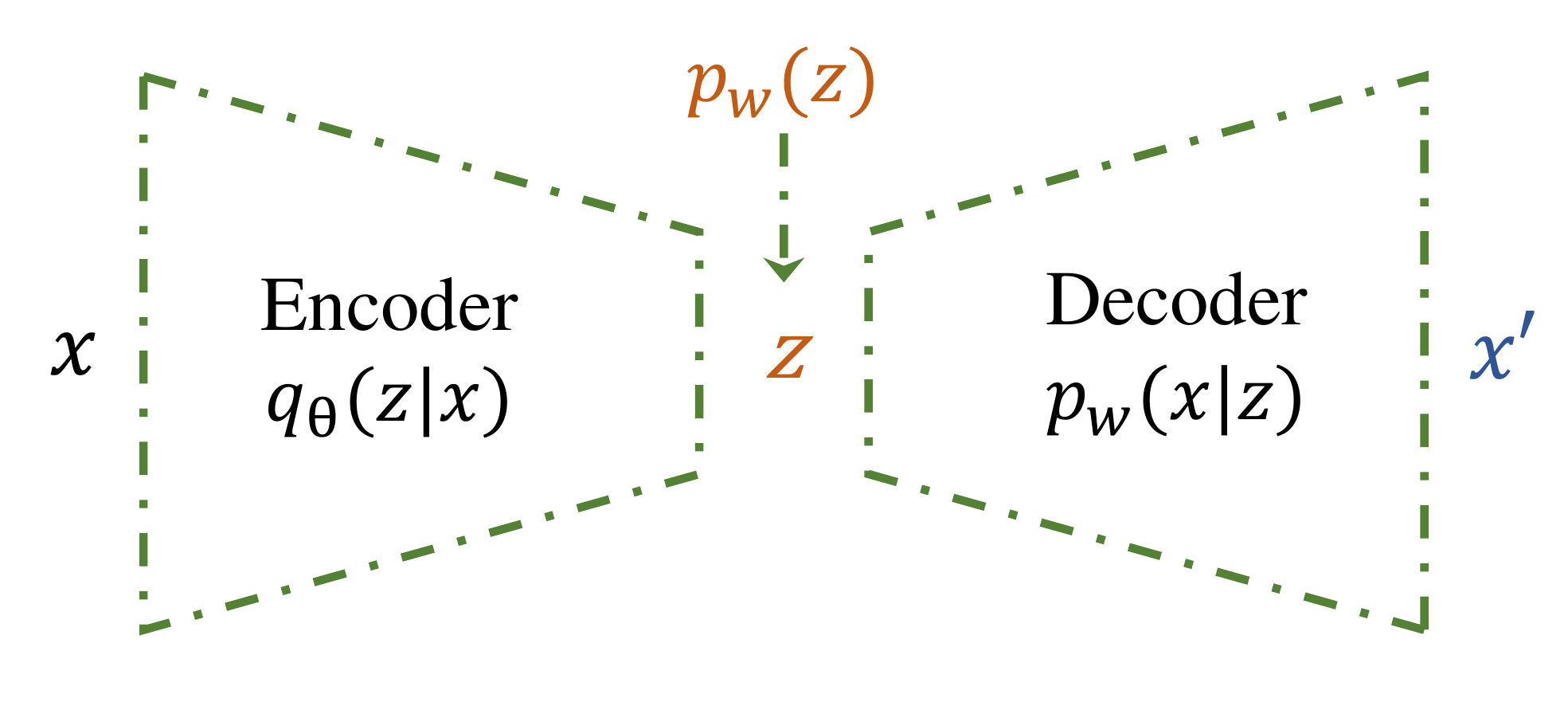} 
    \label{fig:VAE}   
    }     
    \caption{In (a), the encoder and the decoder are denoted by $f(\cdot)$ and $g(\cdot)$, respectively. The notation $\bm{z}$ refers to the latent feature and the notation $\bm{x}'$ refers to the reconstructed data. In (b), both the probabilistic encoder $q_{{\vect{\theta}}}(\vect{z} | \vect{x})$ and the probabilistic decoder $p_{\vect{w}} (\vect{x}| \vect{z})$ are Bayesian networks. The model takes data from $\vect x$-space and passes forward through the inference model to $\vect z$-space, then by a generative model $p_{\bm{w}}(\vect{x},\vect{z})$, the variables $\vect {x'}$ are reconstructed.
    Images adapted from  Ref.~\cite{du2021exploring}.}
    \end{figure}

A typical autoencoder \cite{bourlard1988auto} contains an encoder that aims to compress the input data into a representation, called the bottleneck, and a decoder that seeks to decode the bottleneck to reconstruct the input. The autoencoder is one of the most prominent generative models in the context of unsupervised machine learning  \cite{goodfellow2016deep,lecun2015deep}. It has been widely applied in dimension reduction \cite{wang2016auto,baldi2012autoencoders}, image restoration \cite{vincent2008extracting,zhang2021exposure}, anomaly detection \cite{zhou2017anomaly}, information retrieval \cite{salakhutdinov2009semantic}, and genomics \cite{eraslan2019deep}. As depicted in Fig.~\ref{fig:AE}, given the input data $\vect{x}\in \mathbb{R}^{d_x}$, the encoder $f$ maps it to the hidden units $\vect{z} \in \mathbb{R}^{d_y}$, i.e., the bottleneck with a reduced dimension with $ d_y< d_x$. The mathematical expression is $\vect{z}=f(\vect{W}\cdot\vect{x}+\vect{b})$, where $\vect{W}$ is a $d_y\times d_x$ weight matrix and $\vect{b}$ is the local bias term.  Then the decoder $g$ reconstructs the output $\vect{x'}\in \mathbb{R}^{d_y}$ from the latent space representation $\vect{z}$, i.e.,  $\vect{x'}=g(\vect{W'}\cdot \vect{z}+\vect{b'})$. Here, $\vect{W'}$ is a $d_x\times d_y$ matrix and $\vect{b'}$ is a local bias term.  Commonly, the construction of $f$ and $g$ are realized by MLPs consisting of a bunch of linear projections and non-linear activation functions, like sigmoid and tanh. The optimization of autoencoders aims to reduce the reconstruction error between the input $\vect{x}$ and output $\vect{x'}$, where the squared error and the Kullback-Leibler (KL) divergence \cite{kullback1951information} cost functions are broadly used. The reduction in the dimensionality of the latent space enables the autoencoder to extract useful features and abandon the unimportant ones, in the same vein as principal component analysis (PCA) \cite{tipping1999probabilistic}.
Another typical generative model, the variational autoencoder (VAE) proposed by \citet{Kingma2014}, leverages the variational Bayesian algorithm to learn the joint probability distribution $q(\vect{z},\vect{x})$ for a directed graph model with the latent variable $\vect{z}$ and observable variable $\vect{x}$. The marginal distribution $ p_{\vect{w}}(\vect{x})=\int p_{\vect{w}}(\vect{x}, \vect{z})d \vect{z}$ is intractable generally as there exists no analytical solution or efficient estimator to compute the integral, and this leads to the intractable true posterior distribution $p_{\vect{w}}(\vect{z} | \vect{x})=p_{\vect{w}} (\vect{x}, \vect{z})/ {p_{\vect{w}}(\vect{x})}$.
The intractability can be efficiently mitigated by introducing an inference (or a recognition) model $q_{{\vect\theta}}(\vect{z} | \vect{x})$ with parameters $\vect\theta$, i.e., the probabilistic encoder, which approximates the true posterior distribution. The log-likelihood function of a VAE model is 
\begin{equation}
    \log p_{\vect{w}}(\vect{x}) =
\mathcal{C}_{\vect{\theta},{\vect{w}}}(\vect{x})+D_{K L}\left(q_{{\vect\theta}}(\vect{z} | \vect{x}) \| p_{\vect{w}}(\vect{z} | \vect{x})\right),
\end{equation}
where the term $\mathcal{C}_{\vect{\theta},\vect{w}}(\vect{x}) = \mathbb{E}_{q_{{\vect{\theta}}}(\vect{z} | \vect{x})} \left[ \log \left[{p_{\vect{w}}(\vect{x}, \vect{z})} / {q_{{\vect{\theta}}}(\vect{z} | \vect{x})}\right]\right] $ is called the evidence lower bound (ELBO), and the second term is the KL divergence between the true posterior and the approximate one, which is equal to $\mathbb{E}_{q_{{\vect\theta}}(\vect{z} | \vect{x})}\left[\log \left[{q_{{\vect\theta}}(\vect{z} | \vect{x})}/{p_{\vect{w}}(\vect{z} | \vect{x})}\right]\right]$. Since the KL divergence  is always non-negative, the inequality $\log p_{\vect\theta}(\vect{x}) \geq \mathcal{C}_{\vect{\theta},\vect{w}}(\vect{x})$  holds. By maximizing the ELBO, we can maximize the log-likelihood of the VAE model and concurrently minimize the discrepancy between the two posteriors~\cite{MAL-056}. The optimization routine requires the ELBO to be updated with the gradient information of both $\vect\theta$ and $\vect{w}$. However, computing the gradient $\nabla_{{\vect\theta}} \mathcal{C}_{\vect{\theta},\vect{w}}(\vect{x})$ is not as easy as computing the gradient $\nabla_{\vect{w}} \mathcal{C}_{\vect{\theta},\vect{w}}(\vect{x})$. By applying a reparameterization trick \cite{Kingma2014,MAL-056,rezende2014stochastic}, the latent variable $\vect{z}$ can be expressed as a deterministic and differentiable function $g$, where $\vect{z}=g_{\vect\theta}(\vect\epsilon,\vect{x})$, and the randomness in $\vect z$ is transferred to the auxiliary variable $\vect\epsilon$. This allows one to use the Monte Carlo (MC) estimation and direct differentiation. Two versions of the ELBO estimator are workable,
\begin{equation}\label{vae-a}
    \widetilde{\mathcal{C}}^{A}_{\vect{w}, \vect\theta}(\vect{x})=\frac{1}{L} \sum_{l=1}^{L} \log p_{\vect{w}}(\vect{x}, \vect{z}^{(l)})-\log q_{{\vect\theta}}(\vect{z}^{(l)} | \vect{x})
\end{equation}
and
\begin{equation}\label{vae-b}
\widetilde{\mathcal{C}}^{B}_{\vect\theta, \vect\phi}(\vect{x})=-D_{K L}(q_{{\vect\theta}}({\vect{z}} |\vect{x})|| p_{\vect{w}}(\vect{z}))+\frac{1}{L} \sum_{l=1}^{L}\log p_{\vect{w}}(\vect{x} |\vect{z}^{(l)}),
\end{equation}
where the auxiliary variable $\vect\epsilon^{(l)} $ follows a distribution $ p(\vect\epsilon)$, e.g., the normal distribution or uniform distribution. These estimators are referred to as the stochastic gradient variational Bayes (SGVB) estimators \cite{Kingma2014} and $\widetilde{\mathcal{C}}^{B}_{\vect{w}, \vect\theta}(\vect{x})$ generally has a lower variance than the generic one $\widetilde{\mathcal{C}}^{A}_{\vect{w}, \vect\theta}(\vect{x})$. Moreover, one can see the autoencoder structure implied by Eqn.~\eqref{vae-b}. As illustrated in Fig.~\ref{fig:VAE}, the first KL divergence term measures the encoding error of the approximate posterior $q_{{\vect\theta}}(\vect{z} | \vect{x})$  and the prior $p_{\vect{w}}(\vect{z})$, while the second term represents the reconstruction from the latent $\vect{z}$-space to the observed $\vect{x}$-space, through a probabilistic decoder $p_{\vect{w}}(\vect{x}|\vect{z})$. The encoder $q_{{\vect\theta}}(\vect{z} | \vect{x})$ could be comprised of various distributions \cite{Kingma2014}, such as exponential, uniform, triangular, Gaussian, and chi-Squared, etc. And different distributions usually correspond to specific choices of differentiable $g_{\vect\theta}(\vect{\epsilon}, \vect{x})$, as well as the choices of the auxiliary variable $\bm\epsilon$ . The probabilistic decoder could be Gaussian or Bernoulli MLP according to the type of data.

In recent years, many variants of VAE have been developed, and we refer to some important examples for interested readers. Ref.~\cite{DBLP:journals/corr/BurdaGS15}  proposed an importance weighted autoencoder (IWAE) which increases the flexibility in approximating the true posterior distribution. By employing a GAN substructure in the inference stage, the adversarial autoencoders (AAE) \cite{makhzani2015adversarial} can match the aggregated posterior to an arbitrary prior distribution. The InfoVAE \cite{10.1609/aaai.v33i01.33015885} and $\beta$-VAE \cite{higgins2016beta} introduce regularizers into their ELBOs, which encourages disentanglement learning.
Treating the discrete latent variables as vectors, the vector quantised VAE (VQ-VAE) \cite{van2017neural} could avoid the posterior collapse issue in training. Beyond learning in an unsupervised manner, some variants have been applied to process labeled data \cite{sohn2015learning,makhzani2015adversarial,kingma2014semi}. The VAE model and its variants are broadly applied to image generation \cite{gregor2015draw,kulkarni2015deep}, natural language processing \cite{bowman-etal-2016-generating,kusner2017grammar}, and molecular design \cite{gomez2018automatic}, etc. For some recent developments of VAE models, refer to Refs.~\cite{dai2019diagnosing,kumar2020regularized,vahdat2020nvae}.

\subsection{Quantum mechanics and quantum computing}
In this subsection, we introduce the background knowledge and notations of quantum mechanics, quantum computing, and quantum complexity theory. Refer to \cite{nielsen_chuang_2010} for details.

\subsubsection{Quantum mechanics}

Quantum mechanics provides a conceptual and mathematical framework for describing the physical world. In general, any isolated physical system can be completely represented by a normalized quantum state $|\psi\rangle$, which is a complex 
unit vector in the system's state space (Hilbert space) $\mathcal{H}$. The evolution of such a quantum system can be described by a unitary operator $U$, with $|\psi'\rangle=U|\psi\rangle$ denoting the state after the evolution. In particular, the time evolution of a closed quantum system is characterized by the Schr\"odinger equation,
\begin{equation}
i \hbar \frac{d|\psi(t)\rangle}{dt} =H|\psi(t)\rangle,
\end{equation}
where $\hbar$ is Planck's constant, $|\psi(t)\rangle$ is the state at time $t$ and $H$ denotes the Hamiltonian of the quantum system. Given the state at some initial time $t=0$, if $H$ is independent of time, the state of the system can be derived as $|\psi(t)\rangle=e^{-iHt/\hbar}|\psi(0)\rangle$, with $U(t)=e^{-iHt/\hbar}$ named as the unitary time evolution operator.

The density matrix provides an alternative formulation for describing quantum systems in mixed states. Suppose a quantum system is in state $|\psi_i\rangle$ with  probability $p_i$, the density operator for this system is defined as $\rho=\sum_i p_i |\psi_i\rangle \langle \psi_i |$, where $\langle\psi_i|$ is the conjugate transpose of $|\psi_i\rangle$ and $\rho$ is a normalized, positive semi-definite Hermitian operator. The evolution of a mixed state $\rho$ is accordingly described as $\rho'=U\rho U^\dagger$.

Information contained within a quantum system can be extracted through quantum measurements. The most typical one is the projective measurement, which is described by an orthonormal basis $\{|i\rangle\}$. Performing this measurement on a quantum system $\rho$ will collapse the system to state $|i\rangle$ with a probability of $\langle i|\rho|i\rangle$.
Physical quantities such as position, momentum, and energy can be described by observables $O$, which act on the Hilbert space of the system. Since the observable is Hermitian, this implies, from the spectral decomposition theorem, that $O=\sum_i \lambda_i P_i$, where $P_{i}=|i\rangle \langle i|$ is the projector onto the eigenspace of $O$ with a real eigenvalue of $\lambda_i$. Note that $|i\rangle$ is the corresponding eigenvector and $\lambda_i$ also indicates the possible outcomes of the observable $O$. The expectation value of $O$ is quantified by $\langle O \rangle_\rho=\tr(\rho O)$.
Generally, a wide class of observables are formed by the Pauli matrices
\begin{equation}
\sigma_{x} \equiv\left(\begin{array}{rr}
0 & 1 \\
1 & 0
\end{array}\right),
\quad
\sigma_{y} \equiv\left(\begin{array}{rr}
0 & -i \\
i & 0
\end{array}\right),
\quad
\sigma_{z} \equiv\left(\begin{array}{rr}
1 & 0 \\
0 & -1
\end{array}\right).\nonumber
\end{equation}
And the eigenstates of $\sigma_{z}$ are denoted as
\begin{equation}
|0\rangle \equiv \begin{pmatrix} 1 \\ 0 \end{pmatrix} \ ,\ 
|1\rangle \equiv \begin{pmatrix} 0 \\ 1 \end{pmatrix} \ .
\end{equation}

\begin{table}[htb]
	\centering
	\renewcommand\arraystretch{1.2}
	\caption{Basic quantum gates and their mathematical expressions.}
	\label{tab:Quantum gate}
	\footnotesize
	\includegraphics[width=\linewidth]{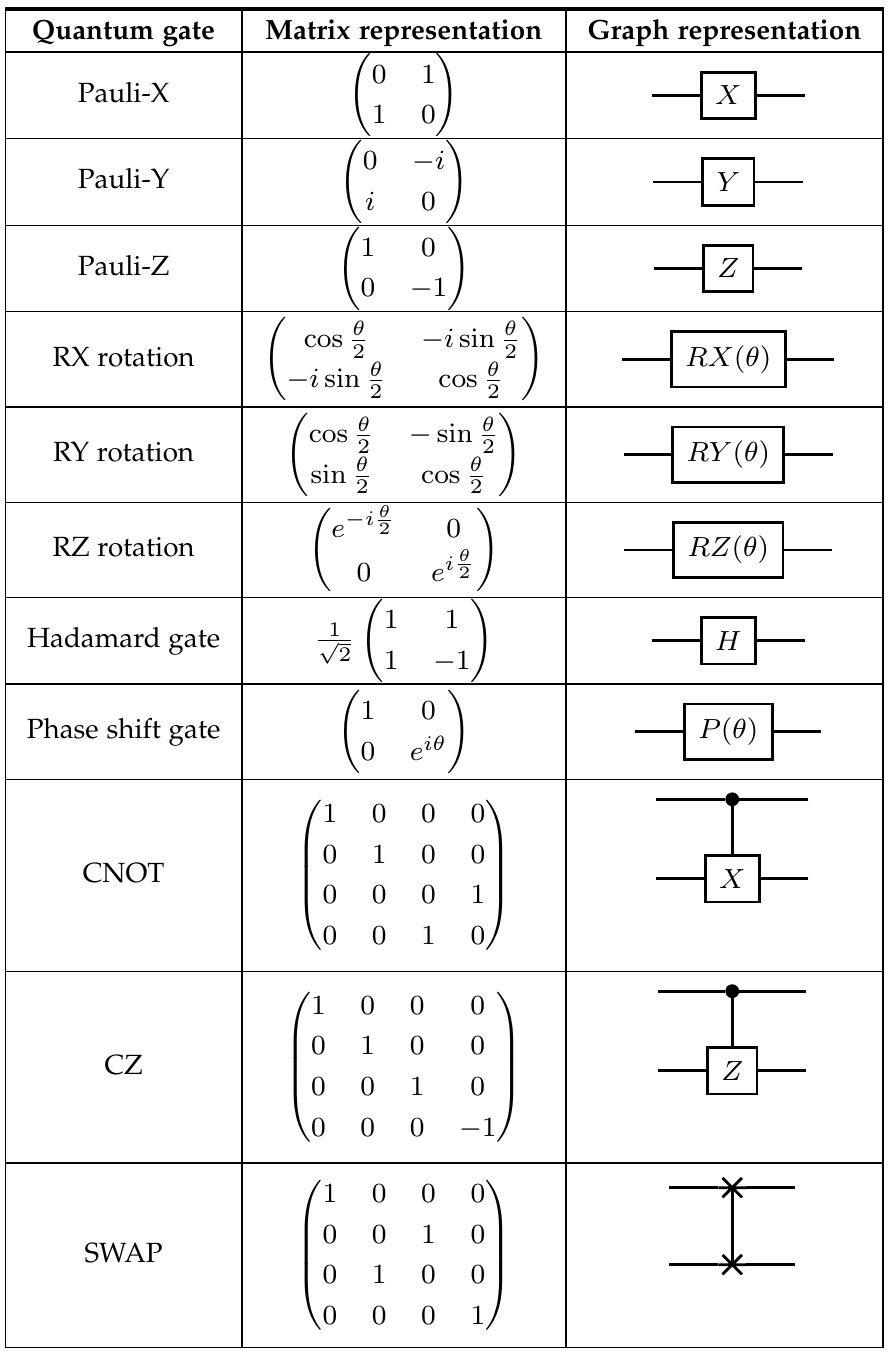}
\end{table}
\subsubsection{Quantum computing}
Quantum computing is an inter-discipline of quantum physics and computation theory that  has recently gained popularity due to the possibility for it to  define the next generation of computers. The concept of quantum computing stems from the realization that simulating quantum evolution is infeasible for classical computers \cite{feynman2018simulating}, since the computational resources required for classical simulation scale exponentially.
Earlier work in the 1980s \cite{benioff1980computer} brought forward a quantum mechanical model of the Turing machine, and promoted research on quantum computing.  Although large fault-tolerant quantum computers are still absent, intermediate-scale quantum devices \cite{preskill2018quantum} which can handle practical applications will arise in the foreseeable future.

Analogous to the way that a classical computer stores information in binary classical bits, quantum computers adopt quantum bits, referred to as \textit{qubits}, as their basic working units.  A qubit is physically implemented by a two-level quantum system, described by a two-dimensional complex vector in the Hilbert space $\mathcal{H}$. Following the Dirac notation (a.k.a, bra–ket notation), a general single-qubit system can be described as  
\begin{equation}
|\psi\rangle:=\alpha|0\rangle+\beta|1\rangle = \begin{pmatrix} \alpha \\ \beta \end{pmatrix} ,
\end{equation}
where $\alpha$ and $\beta$ are complex numbers with the normalization condition that $|\alpha|^{2}+|\beta|^{2}=1$.
Several qubits can be integrated into a multi-qubit quantum system via the tensor product. For example, an $n$-qubits system with each qubit in state $|0\rangle$ can be expressed as $|0\rangle^{\otimes n} \equiv \underbrace{\ket{0 \cdots 0}}_{n}=[1, \underbrace{0,...,0}_{2^n -1}]^{\top}$.

Quantum computers are built from quantum circuits, which consist of wires and quantum gates to transmit and manipulate quantum information. As basic building blocks of quantum circuits, quantum gates manipulate the states of the qubits, affecting the evolution of the overall quantum system. Generally, a unitary evolution is carried out via a series of elementary quantum gates, which are formed by single-qubit and two-qubit gates. Typical elementary quantum gates include the Pauli gates, which can be represented by the Pauli matrices $\sigma_x$, $\sigma_y$ and $\sigma_z$, the single-qubit rotation gates ($RX$, $RY$, $RZ$), the Hadamard gates $H$, and the CNOT gates. Some of the commonly used quantum gates with their matrix and graph representation are summarized in Table.~\ref{tab:Quantum gate}. Note that the single-qubit rotation gates are defined as $RX(\theta) = \exp(-i\sigma_x\theta/2)$, $RY(\theta) = \exp(-i\sigma_y\theta/2)$ and $RZ(\theta) = \exp(-i\sigma_z\theta/2)$. In addition, for the sake of brevity, a Pauli gate with a subscript describes a gate operated on an $n$-qubit system where only one particular qubit is operated. Take $\sigma^{x}_i$ for example,  
\begin{equation*}
  \sigma_{i}^{x} \equiv \overbrace{I \otimes \cdots \otimes I}^{i-1} \otimes \sigma_{x} \otimes \overbrace{I \otimes \cdots \otimes I}^{n-i}.
\end{equation*}
In fact, the single-qubit rotation gates, together with the phase shift gate $P(\phi)$ and the CNOT gates, form a universal gate set. That is, any unitary operation $U$ of an arbitrary dimension can be implemented as a finite sequence of the gates from this set.

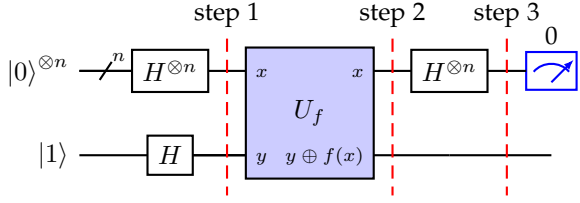
\begin{figure}
\centering
 \begin{quantikz}
\lstick{$\ket{0}^{\otimes n}$} & [2mm] \gate{H^{\otimes n}} \qwbundle{n} \slice{step 1} & \gate[wires=2, style={fill=blue!20}][1.7cm]{U_f} \gateinput{$x$} \gateoutput{$x$} \slice{step 2}  & \gate{H^{\otimes n}} \slice{step 3} & \meter[draw=blue]{0} \\
\lstick{$\ket{1}$}  & \gate{H} & \qw\gateinput{$y$}\gateoutput{$y\oplus f(x)$} & \qw & \qw 
\end{quantikz}
        \caption{The implementation of a quantum circuit for the Deutsch-Jozsa Algorithm.}
    \label{fig:QC_DJ}
\end{figure}

Next, we explain how quantum circuits can be used to complete the Deutsch-Jozsa algorithm, which attains an exponential speedup over the best classical algorithms. The problem step is defined as follows.  Given a boolean function $f: \{0,1\}^n \rightarrow \{0,1\}$, we are promised that $f$ is either a constant function or a balanced function, where a balanced function means of $2^n$ distinct inputs in $\{0,1\}^n$, the function value of half of them is 1 and that of the other half is 0. In quantum computing models, the information of a given function can be encoded into an oracle, which is essentially a black box quantum computer. As shown in Fig.~\ref{fig:QC_DJ}, the Deutsch-Jozsa algorithm can determine whether $f$ is constant or balanced within only one query of the oracle. The algorithm employs $n+1$ qubits and the initial state is $|0\rangle^{\otimes n}|1\rangle$. After the Hadamard transformation in step 1, the state evolves as $\left( 1/\sqrt{2^{n+1}} \right) \sum_{x=0}^{2^{n}-1}|x\rangle(|0\rangle-|1\rangle)$. The oracle $U_f$ which maps state $|x\rangle|y\rangle$ to $|x\rangle|y \oplus f(x)\rangle$ is applied at step 2.
After that, the state becomes $\left( 1/\sqrt{2^{n+1}} \right) \sum_{x=0}^{2^{n}-1} (-1)^{f(x)} |x\rangle(|0\rangle-|1\rangle)$.
At step 3, through applying Hadamard gates to the first $n$ qubits, the final state is
\begin{align*}
\frac{1}{2^{n}} \sum_{y=0}^{2^{n}-1}\left[\sum_{x=0}^{2^{n}-1}(-1)^{f(x)}(-1)^{x \cdot y}\right]|y\rangle,
\end{align*}
where $x \cdot y =  x_0y_0 \oplus \cdots \oplus x_{n-1}y_{n-1}$ and $\oplus$ is addition modulo 2.
The probability of measuring $|0\rangle^{\otimes n}$ is $\left|\frac{1}{2^{n}} \sum_{x=0}^{2^{n}-1}(-1)^{f(x)}\right|^{2}$, which is 1 if $f$ is constant and 0 if balanced.
Therefore, by a single measurement of the result, we can determine whether $f$ is constant or balanced with certainty.

\subsubsection{Quantum complexity theory}

\begin{figure}[t]
    \centering
    \includegraphics[width=0.2\textwidth]{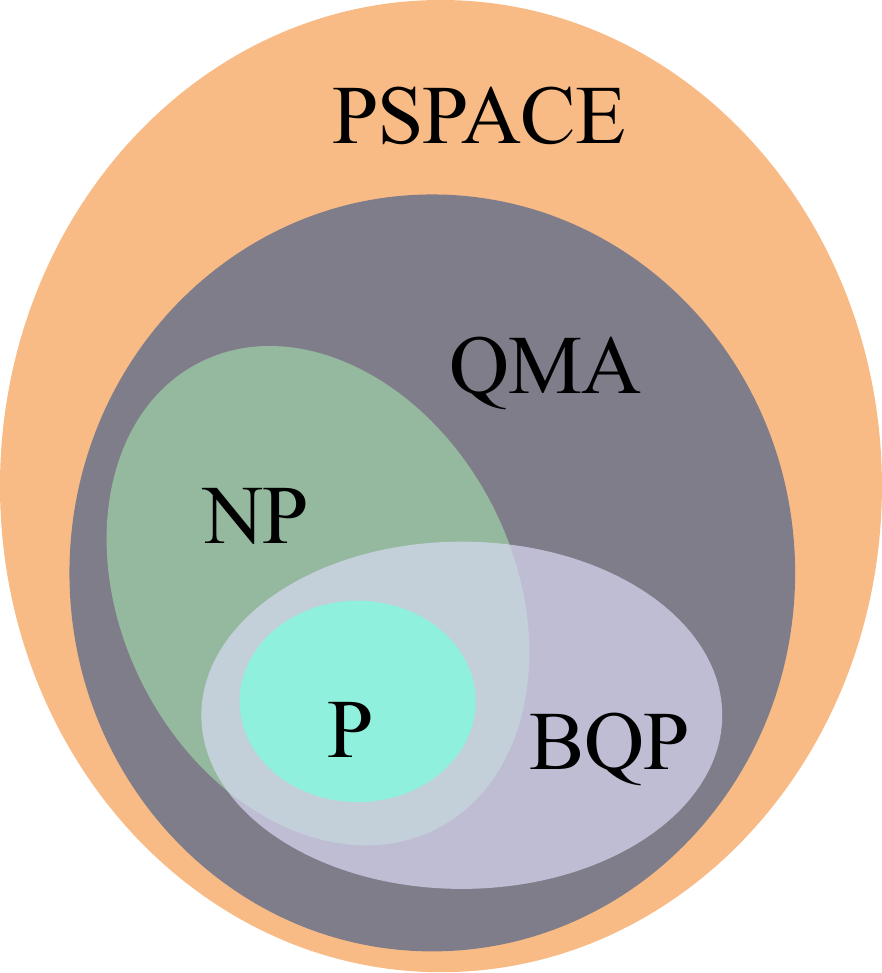}
    \caption{A widely recognized relationship between some complexity classes. P: classically solvable in polynomial time; NP: classically verifiable in polynomial time; BQP: solvable in polynomial time on a quantum computer; QMA: verifiable in polynomial time on a quantum computer; PSPACE: classically solvable in polynomial space.
    } 
    \label{fig:complexity_classes} 
\end{figure}

Quantum complexity theory is a subfield of computational complexity theory.
Theoretical works~\cite{bremner2011classical,bremner2016average,bremner2017achieving,farhi2016quantum} have proven the advantage of quantum computing over a classical computer on some tasks from the perspective of computational complexity theory.
Although there are many unsolved problems in computational complexity theory (e.g., a well-known open problem is whether P is equal to NP), the relation of some frequently used complexity classes shown in Fig.~\ref{fig:complexity_classes} is widely believed.  Specifically, bounded-error quantum polynomial time (BQP) contains decision problems that are solvable by a quantum computer in polynomial time. While relationships between NP and BQP have not been mathematically proven, some problems are solvable in polynomial time by quantum algorithms and only verifiable in polynomial time classically, e.g., integer factorization. Shor's algorithm \cite{shor1999polynomial} solves the integer factorization problem in polynomial time which is exponentially faster than the best known classical algorithm. This quantum algorithm poses a real threat to the current cryptosystem.
Grover's algorithm \cite{grover1996fast}, also known as the quantum search algorithm, searches an unstructured database with $N$ entries using only $\sqrt{N}$ queries. Amplitude amplification is a generalization of Grover's algorithm and the application of amplitude amplification usually leads to quadratic speedups over classical algorithms. However, noisy environments and qubit limitations prevent the implementation of the aforementioned algorithms. To fully utilize the currently available NISQ devices, a different algorithm design philosophy arose in recent years in the form of variational quantum algorithms.

\subsection{Variational quantum algorithms}
\label{subsec:vqa}

Variational quantum algorithms (VQAs) \cite{cerezo2021variational} are a class of quantum-classical hybrid algorithms whose mechanisms resemble deep neural networks. The applications of VQAs include  combinatorial optimization \cite{farhi2014quantum}, quantum chemistry \cite{peruzzo2014variational}, quantum error correction \cite{roffe2019quantum}, and  machine learning  \cite{biamonte2017quantum}. To fully leverage the power of both quantum and classical computers, VQAs implement parameterized quantum circuits on real quantum devices and use classical optimizers to update these trainable parameters. In this way, VQAs can readily adapt to the limitations of near-term quantum devices, which have restricted circuit depth and unavoidable gate noise. The superiority of VQAs stems from the fact that the state space represented by parameterized quantum circuits is unlikely to be efficiently simulated by classical computers. These properties ensure that VQAs are one of the most promising strategies to obtain quantum advantages in the NISQ era~\cite{bharti2022noisy}. 

\begin{figure}
    \centering
    \includegraphics[width=0.47\textwidth]{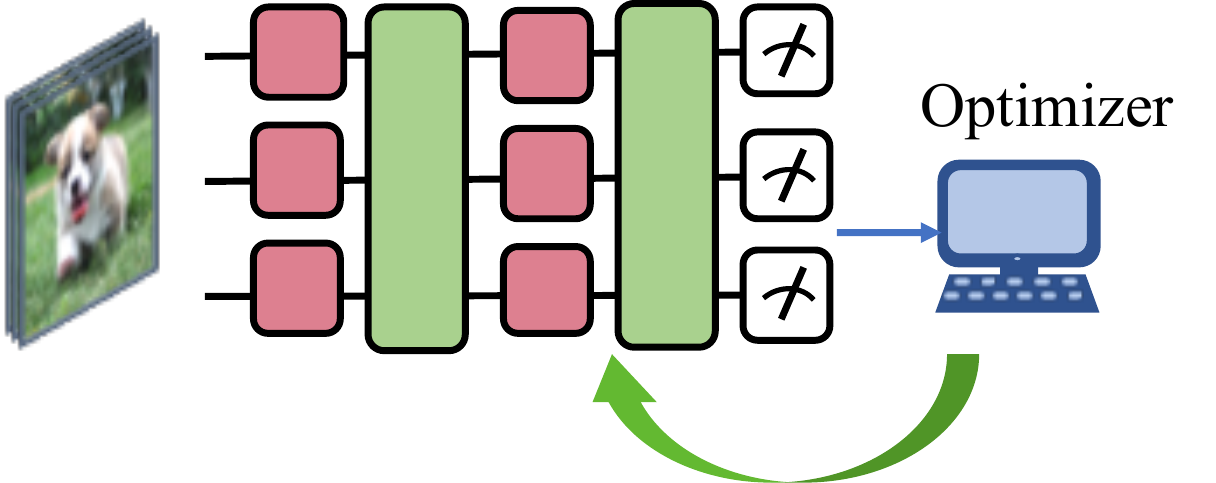}
    \caption{The Schematic of VQAs. The elements of VQAs include the parameterized quantum circuit and the classical optimizer. The red and green boxes refer to the single-qubit and multi-qubit quantum gates, respectively. These gates can either be trainable or fixed.  In the training process, the optimizer continuously updates the trainable parameters to minimize the discrepancy between the quantum circuits' predictions and the true labels.}
    \label{fig:vqas}
\end{figure}
 
The schematic diagram of VQAs is outlined in Fig.~\ref{fig:vqas}. Analogous to NNs, VQAs are composed of three fundamental building blocks: the learning model, the optimizer, and the cost function. The key difference between classical machine learning algorithms and VQAs is in the way the learning model is implemented. Namely, the former adopts deep neural networks while the latter adopts the parameterized quantum circuit (as so-called quantum neural networks (QNNs) in the context of quantum machine learning), which is formed by the parameterized quantum gates and fixed quantum gates. Given a set of training data $\{\bm{x}_i, \bm{y}_i\}_{i=1}^N$, denote the prediction of QNN for the $i$-th example as $\bm{\hat{y}}_i=h(\bm{x}_i, \vect\theta)$ for $\forall i \in [N]$. The optimization of VQAs amounts to finding the optimal parameters $\vect\theta^*$ and minimizing the distance between the true labels and predictions according to a predefined cost function $\mathcal{C}(\cdot, \cdot)$, i.e., $\vect\theta^* =  \arg\min_{\vect\theta}\sum_{i=1}^N \mathcal{C}(\bm{y}_i, h(\bm{x}_i, \vect\theta))$. The possible choice of cost functions contains the mean square loss and the cross-entropy loss in discriminative learning and KL divergence in generative learning. The optimization can be completed via both gradient-free or gradient descent methods. The gradients can be acquired via the parameter shift rule \cite{schuld2019evaluating}. 
 
Next, we elucidate the mathmatical foundation of QNN $h(\bm{x}_i, \vect\theta)$, which is constructed by the initial state, the encoding unitary $U_{\bm{x}}$, the Ansatz $U(\vect\theta)$, and the measurement operator. Given the classical data, which can be the training example or the instance sampled from the prior distribution, QNN encodes it into the quantum state via the encoding unitary $U_{\bm{x}}$. Note that the method used to achieve $U_{\bm{x}}$ is flexible and can be via basis encoding, amplitude encoding, and gate encoding. For instance, when the gate encoding is adopted, the encoding unitary may take the form  $U_{\bm{x}}=\otimes_i RY(\bm{x}_i)$. Another key component in QNN is Ansatz $U(\vect\theta)$. As with deep neural networks, Ansatz obeys a multi-layer structure, i.e., \begin{equation}
    \label{eq:vqa}
    U(\boldsymbol{\theta})=U_{L}\left(\boldsymbol{\theta}_{L}\right) \cdots U_{2}\left(\boldsymbol{\theta}_{2}\right) U_{1}\left(\boldsymbol{\theta}_{1}\right).
\end{equation}
Particularly, the gate arrangement in each layer is diverse. As detailed below, a well-designed Ansatz can dramatically enhance the performance of VQAs for certain learning tasks. Let us denote the initial state as $\ket{\Psi_0}$ and the measurement operators as $\{O_i\}_{i=1}^d$ when the dimension of the prediction is $d$. There are two strategies to implement QNN. The first one only queries the data unitary once, i.e., $h(\bm{x}_i, \vect\theta)$ equals to \[[\tr(O_1U(\vect\theta)U_{\bm{x}}\rho U_{\bm{x}}^{\dagger}U(\vect\theta)^{\dagger}),...,\tr(O_dU(\vect\theta)U_{\bm{x}}\rho U_{\bm{x}}^{\dagger}U(\vect\theta)^{\dagger})].\] 
The second one is using $U_{\bm{x}}$ and $U(\vect\theta)$ multiple times, i.e., the $j$-th component of $h(\bm{x}_i, \vect\theta)$  for $\forall j \in [d]$ becomes 
\[\tr(O_jU(\vect\theta)U_{\bm{x}}...U(\vect\theta)U_{\bm{x}}\rho U_{\bm{x}}^{\dagger}U(\vect\theta)^{\dagger}...U_{\bm{x}}^{\dagger}U(\vect\theta)^{\dagger}).\]

As mentioned above, the design of Ans\"atze is critical in VQAs. This is because some gates are costly to construct from native gates. Furthermore, employing more gates hints at introducing more noise into the quantum system in light of the immaturity of current quantum devices. Therefore, choosing effective Ans\"atze depending on the given learning tasks can alleviate these issues. Based on whether the problem Hamiltonian is available, Ans\"atze of VQAs can be roughly classified into two categories, i.e., \emph{Hamiltonian-informed} Ans\"atze and \emph{Hamiltonian-agnostic} Ans\"atze. 

\textit{\underline{Hamiltonian-informed Ansatz}.}
For most quantum physics problems where the Hamiltonian is provided, we could tailor the Ansatz by using the information of the Hamiltonian.
Specifically, each unitary gate is the time evolution of a certain Hamiltonian $H$ in the form of $U(t) = e^{- i t H}$. Thus, we could control the size of the Ansatz by only keeping gates that are relevant to the problem Hamiltonian. For a Hamiltonian that is hard to simulate directly but can be expressed as the linear sum of easily-implemented Hamiltonians $H = \Sigma_k a_k h_k$, the Suzuki-Trotter expansion is commonly used to approximate the corresponding time evolution, i.e.,
\begin{equation}
e^{-i H t}=\lim _{m \rightarrow \infty}\left(\prod_{k} e^{-i \frac{a_{k} h_k}{m} t}\right)^{m}.
\end{equation}

A concrete example of the Hamiltonian-informed Ansatz is the quantum alternating operator Ansatz (QAOA) \cite{hadfield2019quantum}, which is exploited to solve combinatorial optimization problems. Concisely, the target graph is encoded into the problem Hamiltonian $H_P$, and the mixing Hamiltonian $H_M$ is defined as $H_M=\Sigma \sigma_x^{i}$.
The mathematical expression of QAOA is $U(\gamma, \boldsymbol{\beta})=\prod_{l=1}^{p} e^{-i \beta_{l} H_{M}} e^{-i \gamma_{l} H_{P}}$. The performance of QAOA improves monotonically with increasing $p$. Besides QAOA,  Hamiltonian-informed Ans\"atze are also widely applied in quantum chemistry problems, such as the unitary coupled-cluster Ansatz \cite{taube2006new} and the factorized unitary coupled-cluster \cite{chen2021quantum}.

\begin{figure}[t]
    \centering
    \includegraphics[width=0.34\textwidth]{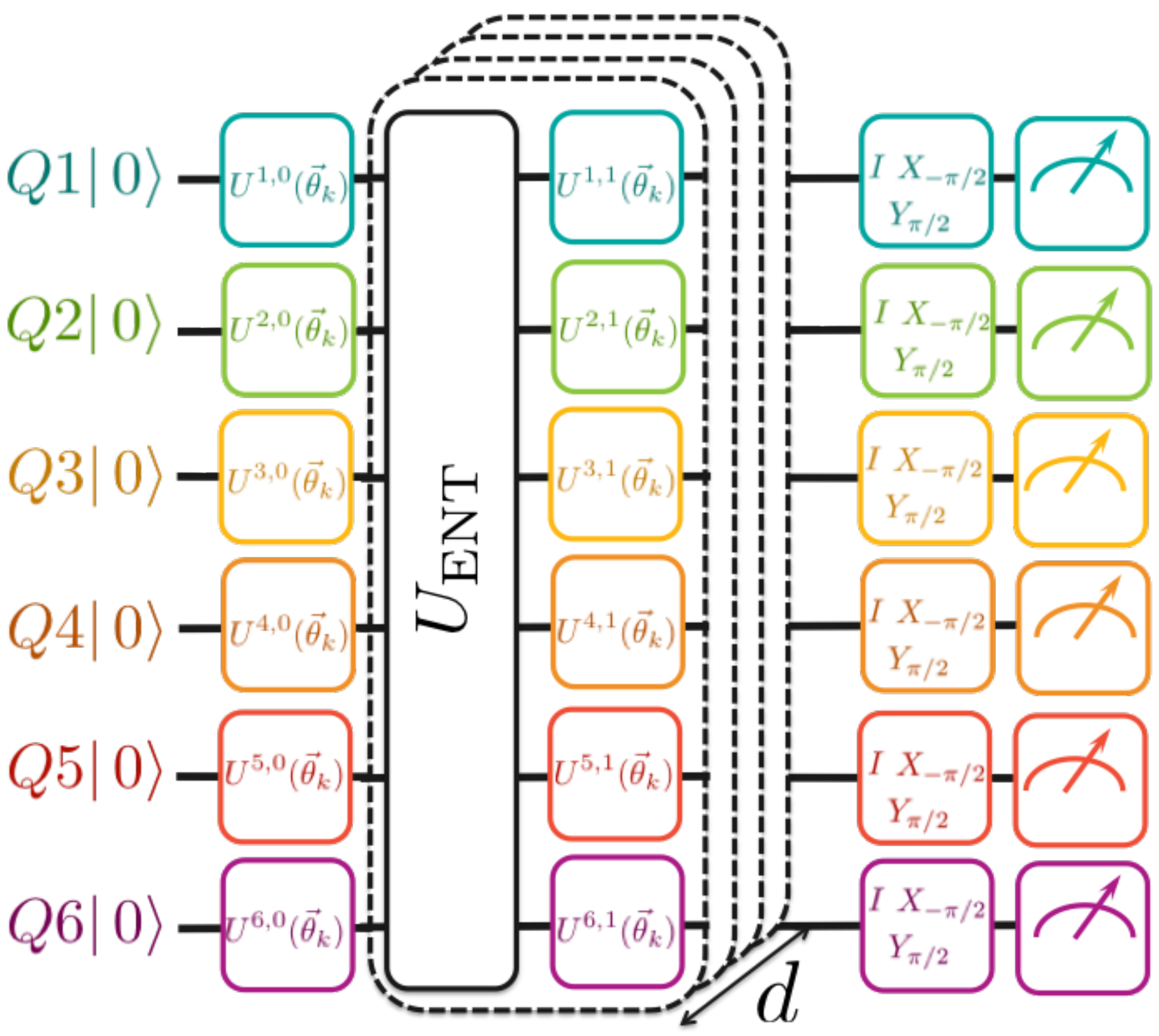}
    \caption{A possible implementation of the hardware-efficient Ansatz. It is composed of the single-qubit rotation gates and multi-qubit gates as native gates. $U_{\text{ENT}}$ represents entangling unitary operations.
    Image taken from Ref.~\cite{kandala2017hardware}.
    }
    \label{fig:hardware-efficient}
\end{figure}

\textit{\underline{Hamiltonian-agnostic Ansatz}.}
Hamiltonian-informed Ans\"atze may not ensure good  VQA performance due to challenges including the constrained qubit connectivity, the restricted gate sets, the imperfect gate fidelities, and limited coherence times of current quantum devices. Moreover, for most conventional machine learning problems, less is known about how to construct the corresponding Hamiltonian. In this scenario, Hamiltonian-agnostic Ans\"atze are broadly employed in VQAs instead of Hamiltonian-informed Ans\"atze.

The most commonly used Hamiltonian-agnostic Ansatz is the hardware-efficient Ansatz \cite{kandala2017hardware} (see Fig.~\ref{fig:hardware-efficient}).
A hardware-efficient Ansatz consists of single-qubit rotation gates and multi-qubit gates, e.g., CNOT or CZ gates. The connectivity of the multi-qubit gates can readily adapt to the topology of the quantum hardware, which suppresses the influence of noise.
Two other classes of Hamiltonian-agnostic Ans\"atze are tensor-network Ans\"atze and neural-network Ans\"atze. The former is inspired by tensor network techniques, where the architecture of the Ansatz mimics classical tensor networks such as the matrix product state (MPS)
\cite{verstraete2008matrix}, the tree tensor network (TTN)
\cite{shi2006classical}, and the multi-scale entanglement renormalization Ansatz (MERA) \cite{vidal2007entanglement}. Similarly, the latter is inspired by deep neural networks, where the architecture of the Ansatz mimics classical neural networks such as convolutional neural networks \cite{cong2019quantum}, recurrent neural networks \cite{bausch2020recurrent}, and graph neural networks \cite{verdon2019quantum}.

\begin{table*}[htb]
	\centering
	\renewcommand\arraystretch{1.2}
	\caption{Representative works of quantum generative learning models from both theoretical and experimental perspectives.}
	\label{tab:RepresentativeWorks}
	\footnotesize
	\begin{tabular}{p{1.1cm}<{\centering}|p{4.4cm}<{\centering}|p{11.3cm}<{\centering}}
		\Xhline{1.2pt}
		\textbf{Category} &  \textbf{Reference}  & \textbf{Highlights}\\
		\hline
		\multirow{5}{*}{QCBM} & \citet{benedetti2019generative}& Propose the first QCBM protocol and demonstrate its ability on a trapped ion quantum computer. \\
		\cline{2-3}
		& \citet{liu2018differentiable}& Introduce the gradient-based optimizers in QCBM and propose to tailor Ansatz with Chow-Liu tree topology.\\
		\cline{2-3}
		& \citet{zhu2019training}& Implement QCBM \cite{benedetti2019generative}  on a seven-qubit fully programmable trapped ion quantum computer.\\
		\cline{2-3}
		& \citet{du2020expressive}& Utilize Bayes inference with the equipment of ancillary qubits and show stronger expressive power than traditional PQCs.\\
		\cline{2-3}
		& \citet{coyle2020born}& Devise QCIBM based on Ising Hamiltonians and propose to use SHD, SD, as well as the quantum kernel MMD as cost functions.\\
		\hline
		%%%%%%%%%%%%%%%%%%%%%%%%%%%%%%%%%%%%%%%%%%%%%
		\multirow{3}{*}{QGAN} & ~\citet{lloyd2018quantum}   &  Propose four types of QGANs, categorized by whether the learning task and the implementations of generators and discriminators are quantum or not. \\
		\cline{2-3}
		& \citet{dallaire2018quantum}   &  A companion paper with Ref.~\cite{lloyd2018quantum}, which mainly focuses on demonstrating the feasibility of QGAN through numerical simulations.\\
		\cline{2-3}
		& \citet{hu2019quantum}   &  Present the first proof-of-principle experimental demonstration of QGAN in a superconducting quantum circuit. \\
		\cline{2-3}
		& \citet{zoufal2019quantum}   &  Exploit QGAN to learn and load generic probability distributions that are implicitly given by data samples into quantum states.  \\
		\cline{2-3}
		& \citet{huang2021experimental}   &  Devise Patch-GAN and Batch-GAN to generate images on a superconducting quantum computer. \\
		\hline
		%%%%%%%%%%%%%%%%%%%%%%%%%%%%%%%%%%%%%%%%%%%%%
		\multirow{4}{*}{QBM} & \citet{amin2018quantum}& Introduce the  noncommutative nature of quantum Hamiltonian into BMs and perform a bound-based training of fully-visible QBMs and semi-restricted ones.\\
		\cline{2-3}
		& \citet{PhysRevA.96.062327}  &  Propose a state-based training scheme for QBMs, enabling quantum state estimation and the state copy reconstruction.\\
		\cline{2-3}
		& \citet{10.5555/3370185.3370188}  & Demonstrate the effectiveness of QBMs compared with classical RBMs in  RL tasks.\\
		\cline{2-3}
		& \citet{zoufal2021variational}   & Design a variational QBM model implemented by PQCs, which enables the Gibbs state preparation and the exact gradient calculation of cost functions. \\
		\hline
		%%%%%%%%%%%%%%%%%%%%%%%%%%%%%%%%%%%%%%%%%%%%%
		\multirow{5}{*}{QAE} & \citet{romero2017quantum}&  Propose the first QAE to compress particular quantum states from high dimensional Hilbert space to a low one.\\
		\cline{2-3}
		& \citet{wan2017quantum}   & Devise a quantum-neuron-based QAE model and discuss its possibility in discovering  quantum teleportation protocols. \\
		\cline{2-3}
		& \citet{khoshaman2018quantum}   & Design a QVAE model in which the latent space of a classical VAE is represented by a QBM, and the training is realized by minimizing Q-ELBO.\\
		\cline{2-3}
		& \citet{steinbrecher2019quantum}   & Introduce a QONN structure, which is likely to be used in the next-generation of quantum processors, since it integrates many quantum optics features with neural networks.  \\
		\cline{2-3}
		& \citet{pepper2019experimental}   & Experimentally realize a QAE which compresses qutrits to qubits with a photonic device.\\
		\hline
	\end{tabular}
\end{table*}

\section{Quantum generative learning models}
\label{sec:QGLM}

In this section, we systematically review the current progress of quantum generative learning models (QGLMs), which can be interpreted as the quantum extension of classical generative learning models as illustrated in Fig.~\ref{fig:overview}. In each subsection, we first elaborate on the seminar work of each class of QGLMs, followed by demonstrating their variants and potential applications. For comprehension, we summarize representative works of QGLMs in Table~\ref{tab:RepresentativeWorks} and the available source codes of QGLMs in Table~\ref{tab:implementations}.
\begin{table}[H]
	\centering
	\caption{A summary of open source codes for QGLMs.}
	\label{tab:implementations}    
    \begin{tabular}{c|c|c|c|c}
		\Xhline{1.2pt}
          & QCBM & QGAN & QBM & QAE \\
        \hline
        TFQ \cite{broughton2020tensorflow}  & - & \cite{niu2021entangling}  &  - & -  \\
        Pennylane \cite{bergholm2018pennylane} & - &  \cite{li2021quantum}  & - & -  \\
        Qiskit \cite{aleksandrowicz2019qiskit} & - &   \cite{stein2020qugan,zoufal2019quantum}  & - & \cite{achache2020denoising}  \\
        Yao.jl \cite{luo2020yao} & - & \cite{zeng2019learning}  & - & -  \\
        Pyquil \cite{smith2016practical} &  \cite{coyle2020born}  & - & - & -   \\
        Others & \cite{liu2018differentiable} & \cite{chakrabarti2019quantum}  & - & - \\    
        \hline
    \end{tabular}
\end{table}

\subsection{Quantum circuit Born machine}
\label{sec:QCBM}

The quantum circuit Born machine (QCBM) was first proposed by \citet{benedetti2019generative} (which is also referred to as the data-driven quantum circuit learning framework). QCBM is the quantum counterpart of the MLP generative learning model formulated in Section~\ref{subsec:MLP-GLM} in which that MLP is replaced with QNNs \cite{benedetti2019parameterized}. The key concept of QCBM is leveraging QNNs to generate a tunable and discrete probability distribution $p_{\vect\theta}$, which approximates the target  discrete probability distribution $q$. QCBMs inherit the Born rule naturally and can be efficiently implemented on NISQ devices due to the versatility of PQCs. 

\begin{figure}[htbp]
    \centering
    \includegraphics[width=0.48\textwidth]{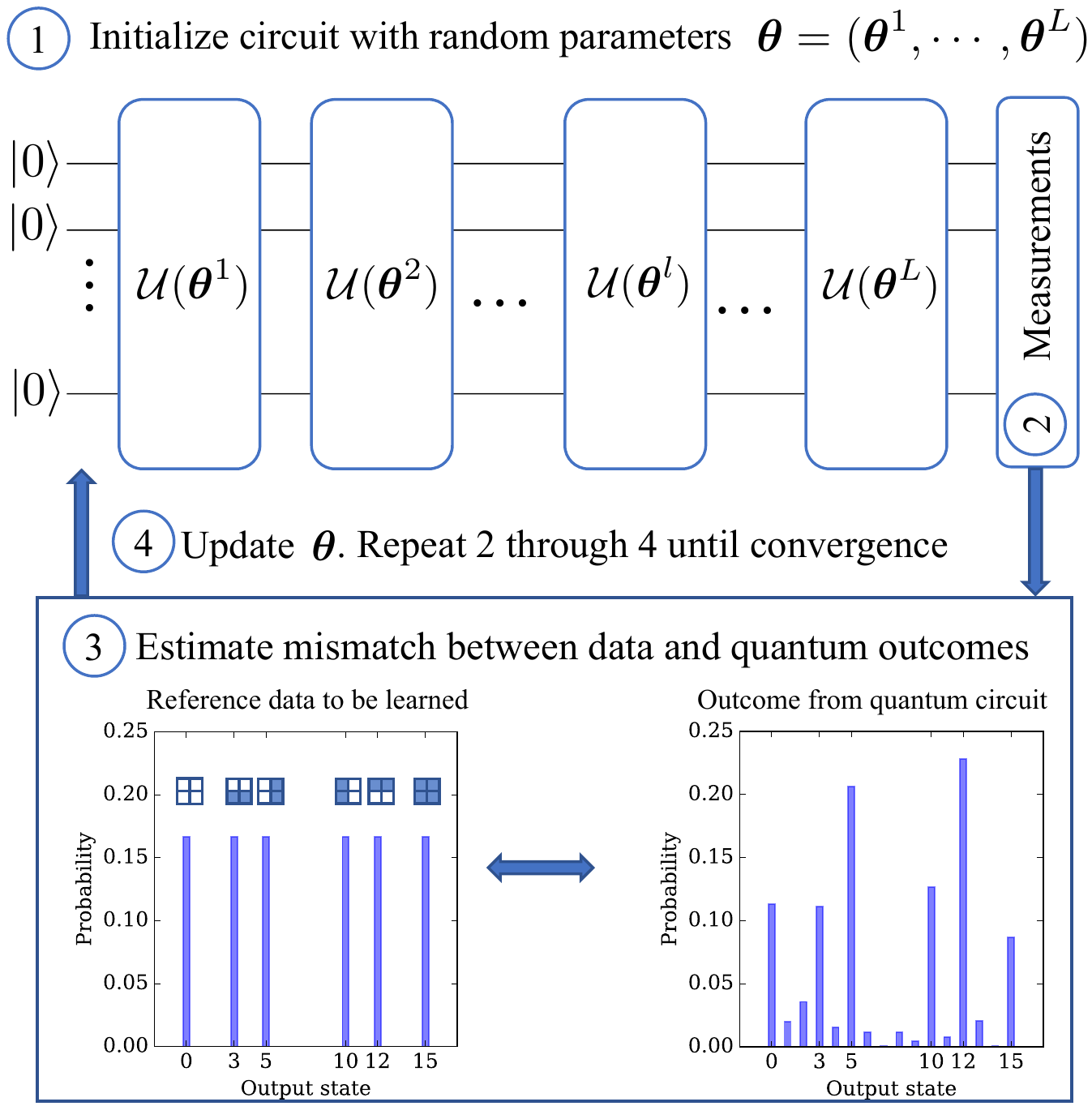}
    \caption{The QCBM learning framework with an illustrative task on Bars And Stripes (BAS) (2,2) image (a $2\times 2$ pixel synthetic dataset image) generation \cite{mackay2003information}. Figure adapted from Ref.~\cite{benedetti2019generative}.}
    \label{fig:QCBM}
\end{figure}

The paradigm of QCBM is illustrated in Fig.~\ref{fig:QCBM}, which is formed by a QNN, a cost function, and a classical optimizer. In Ref.~\cite{benedetti2019generative}, these three components are realized by the hardware-efficient Ansatz \cite{kandala2017hardware}, the KL divergence, and the gradient-free particle swarm optimizer (PSO) \cite{kennedy1995particle}, respectively. More specifically, an initial n-qubit state $|0\rangle^{\otimes n}$ is fed into the hardware-efficient Ansatz $U(\vect\theta)\in \text{SU}(2^n)$ formulated in Section~\ref{subsec:vqa}.  In QCBM, the probability of sampling $\vect{x}\in\{1,2,...,2^n\}$ yields $p_{\vect\theta}(\vect{x})= \tr (\Pi_{\vect{x}} U(\vect\theta) (|0\rangle\langle0|)^{\otimes n} U(\vect\theta)^{\dagger})$,  where $\Pi_{\vect{x}} = |\vect{x} \rangle \langle \vect{x}|$ denotes the projection operator along the computational basis $\vect{x}$. The KL divergence is then adopted to measure the discrepancy between the generated discrete distribution $p_{\vect\theta}$ and the target discrete distribution $q$, i.e., $D_{KL}[q||p_{\vect\theta}]=\sum_{\vect{x}} q(\vect{x}) \log( q(\vect{x})/p_{\vect\theta} (\vect{x}))$. To better approximate $q$, the PSO is employed to continuously update $\vect\theta$, and minimize $D_{KL}[q||p_{\vect\theta}]$ until it converges.

\citet{benedetti2019generative} applied QCBM to complete three learning tasks on both numerical simulators and a trapped-ion quantum computer \cite{debnath2016demonstration}. These learning tasks are BAS (2,2) image generation, 3-qubit Greenberger-Horne-Zeilinger (GHZ) state preparation  \cite{greenberger1990bell,monz201114}, and coherent Boltzmann probability distributed thermal state preparation. The experimental results for the image generation confirmed that the performance of QCBM depends on the topology of the adopted entangling layers in PQCs. That is, when the topology of PQCs accommodates the hardware architectures, QCBM can attain high performance. On the task of GHZ state preparation, the trained QCBM recovered the most efficient and compact protocols on trapped-ion computers \cite{monz201114}.  On the task of thermal state preparation, experimental results indicated that a deeper circuit may achieve better performance when the learning tasks become difficult \cite{choromanska2015loss}.

\subsubsection{Variants}
Since \citet{benedetti2019generative}, extensive studies exploring the enhancement of QCBM capabilities have been conducted. The related literature can be categorized into three classes, i.e., the design of different Ans\"atze, the selection of alternative cost functions, and the employment of powerful optimizers.   

\textit{\underline{Ans\"atze design}.}  Several studies have investigated the possibility of tailoring the hardware-efficient Ansatz to improve the performance of QCBM. In particular, \citet{liu2018differentiable} proposed a variant of hardware-efficient Ansatz whose entangling layers are realized by the Chow-Liu tree structure \cite{chow1968approximating,koller2009probabilistic}. Meanwhile, a differentiable learning method, which is completed by the parameter shift rule, was employed to optimize QCBMs. The simulation results conducted on the BAS (3,3) dataset and the double Gaussian peak model validated the feasibility of their approach. \citet{hamilton2019generative} benchmarked the performance of QCBM implemented on IBMQ\_Melbourne \cite{ibmqcloudmelbourne}. When the number of entangling layers is increased to $2$, a QCBM with sparse connectivity in the entanglement layer attained a better performance than the denser one. In addition, they identified that the optimization may be stuck into local minima for certain configurations of the entanglement layer.  \citet{zhu2019training} experimentally studied the performance of QCBM executed on a customized programmable trapped-ion system \cite{debnath2016demonstration}, and the experimental results showed that the convergence behaviour is closely related to both the optimizer and hardware connectivity. \citet{leyton2021robust} performed a robust test of QCBM on the Rigetti Aspen superconductive quantum processor using a tailored hardware-efficient Ansatz, which is formed by RY gates and CZ gates. The abandonment of RZ gates allows for a reduced number of trainable parameters and thus facilitates accelerated optimization. Following the same routine, \citet{hamilton2021mode} conducted extensive simulations to investigate how different parameterizations (i.e., RY rotation gates versus arbitrary rotation gates) and varied settings for the sparsity of the entangling layers (i.e., periodic closure versus unitary 2-designs \cite{cerezo2021cost}) affect the performance of QCBM. By analyzing the loss landscape, the authors unveiled that although employing RY gates enables fast training, this strategy unavoidably induces rigidity into the circuits. On the other hand, employing arbitrary rotation gates benefits a deeper circuit and generally yields better performance, since it improves the connectivity of the loss landscape. Besides engineering the layout of the hardware-efficient Ansatz, \citet{du2020expressive} enhanced the expressive power of QCBM by introducing ancillary qubits. Numerical simulations verified the effectiveness of their proposal. Recently, \citet{gong2022born} attempted to integrate tensor-network-based Ans\"atze \cite{huggins2019towards} with QCBM. The simulation results indicate that their proposal can attain a comparable generative capability with other QCBMs while require fewer qubit resources and parameters.

Initial attempts have been conducted to combine QCBM with the Hamiltonian-informed Ansatz. Specifically, \citet{coyle2020born} proposed the quantum circuit Ising Born machine (QCIBM). The authors argued that QCIBM could achieve quantum advantages when the employed cost function, i.e., total variation (TV) distance, is less than a small threshold. Experiments on the Regetti platform exhibited the expressive power of QCIBM to learn random instances of target probability distributions. Meanwhile, the authors claimed that QCIBM can be used to perform a type of `weak' quantum circuit compilation. Let us recall that the objective of quantum circuit compilation is to represent an arbitrary unitary $U$ as a sequence of native gates in a quantum device. With the limited quantum resources,  `weak' quantum circuit compilation aims at simulating the output of the target unitary instead of the unitary itself. They used QCIBM with QAOA \cite{farhi2014quantum} to learn the output distribution of an instantaneous quantum polynomial time (IQP) \cite{shepherd2009temporally} circuit.

\textit{\underline{Cost functions}.} Another line of research in the context of QCBM is designing different cost functions. Besides KL divergence applied in Refs.~\cite{benedetti2019generative,liu2018differentiable,zhu2019training,leyton2021robust,alcazar2020classical,zhu2021generative,gong2022born}, the maximum mean discrepancy (MMD) \cite{scholkopf2002learning,gretton2012kernel} has also been widely utilized in Refs.~\cite{liu2018differentiable,du2020expressive,hamilton2019generative,coyle2020born,Hamilton2020ErrormitigatedDC,coyle2021quantum,hamilton2021mode,vcepaite2022continuous,gong2022born} for QCBMs. The mathematical form of MMD is $\text{MMD} [p, q] =\mathbb{E}_{\vect{x},\vect{x}'\sim p}[k(\vect{x},\vect{x}')] 
-2\mathbb{E}_{\vect{x}\sim p,\vect{y}\sim q}[k(\vect{x},\vect{y})]+\mathbb{E}_{\vect{y} ,\vect{y}'\sim q}[k(\vect{y},\vect{y}')]$,
where $k(\vect{x},\vect{y})$ is the kernel function (e.g., Gaussian kernel \cite{liu2018differentiable} or quantum kernel \cite{schuld2019quantum,havlivcek2019supervised} applied in Refs.~\cite{coyle2020born,vcepaite2022continuous}). Other alternative cost functions include the Stein discrepancy (SD) \cite{stein1972bound}, the Sinkhorn divergence (SHD) \cite{ramdas2017wasserstein,cuturi2013sinkhorn}, and the Jensen-Shannon (JS) divergence \cite{leyton2021robust}   derived from KL divergence.  \citet{liu2018differentiable} pointed out that the MMD cost can be more efficiently applied in large-scale systems since KL divergence is usually inaccessible. Furthermore, \citet{coyle2020born} claimed that SHD is superior to MMD in the measure of accuracy and convergence rate. Numerical simulation results certified  the advantages of SHD on certain tasks.

\underline{\textit{Optimizers}.} Both the gradient-free and gradient-based optimizers have been broadly used in QCBM. For gradient-free optimizers, PSO \cite{kennedy1995particle} has been used in Refs.~\cite{benedetti2019generative,zhu2019training}, which enables faster training by using a small amount of memory. The covariance matrix adaptation evolution strategy (CMA-ES) \cite{hansen2006cma}  is capable of optimizing non-linear and non-convex cost functions \cite{articleRIOS}, and it has been exploited in Refs.~\cite{liu2018differentiable,alcazar2020classical} to attain better performance than the simultaneous perturbation stochastic approximation (SPSA) method \cite{spall1998implementation}, but it has nevertheless, underperformed in comparison to Adam when the quantum circuit noise is considered \cite{liu2018differentiable}. Bayesian optimization (BO) \cite{frazier2018tutorial}, as a black-box derivative-free global optimization method, has been explored in Ref.~\cite{zhu2019training}. It constructed a surrogate Bayesian statistical model for the same objective function and queried an acquisition function consisting of the posterior distribution to decide the next sampling direction. The BO method can be applied to problems involving a large number of parameters but the number of queries for evaluating cost functions is limited. Refs.~\cite{kondratyev2021non,coyle2021quantum} adopted Genetic Algorithm (GA) \cite{holland1992genetic}  to optimize QCBMs, which can avoid some local minima in principle. Last, a newly proposed zeroth-order optimization method called ZOOpt toolbox \cite{liu2017zoopt} has been used in Ref.~\cite{leyton2021robust}, which is designed for large-scale systems and is competent at suppressing the influence of noise. It attained the lowest cost value in the robust test in comparison to Adam and the Stochastic Variation of Hill-Climbing type algorithm (SVHC) \cite{leyton2021robust}.

The \emph{gradient-based} optimizers can search local optima efficiently towards deep circuits and large parameter spaces. In this regard, the Adam optimizer has been broadly applied to optimize QCBM models \cite{liu2018differentiable,du2020expressive,coyle2020born,coyle2021quantum,hamilton2019generative,leyton2021robust,gong2022born}.  The Adam optimizer generally requires a modest amount of computational resources and is well-adapted to most differentiable cost functions. The L-BFGS-B (bound limited memory BFGS) is another gradient-based optimizer adopted in Ref.~\cite{liu2018differentiable}. It has a comparable efficiency to L-BFGS while its performance is superior when handling constrained problems \cite{nocedal1999numerical}. Although it is not noise-tolerant, L-BFGS is particularly favored in large systems (e.g. $\geq 1000$ variables) due to its fast convergence rate. \citet{liu2018differentiable} compared the performances of Adam and CMA-ES while treating the L-BFGS-B as a reference with an infinite batch size. The simulation results indicated that CMA-ES is more vulnerable to noise and converges more slowly than Adam.

Although the selection of a suitable optimizer is  typically learning task dependent, it can be concluded that gradient-based methods are capable of dealing with a large-scale QNN and are less sensitive to system noise. Gradient-free optimizers are suitable for a small-scale QNN with fewer parameters \cite{liu2018differentiable} and  highly non-smooth cost functions \cite{kondratyev2021non}.

\begin{figure*}[htbp]
    \centering
    \includegraphics[width=0.8\textwidth]{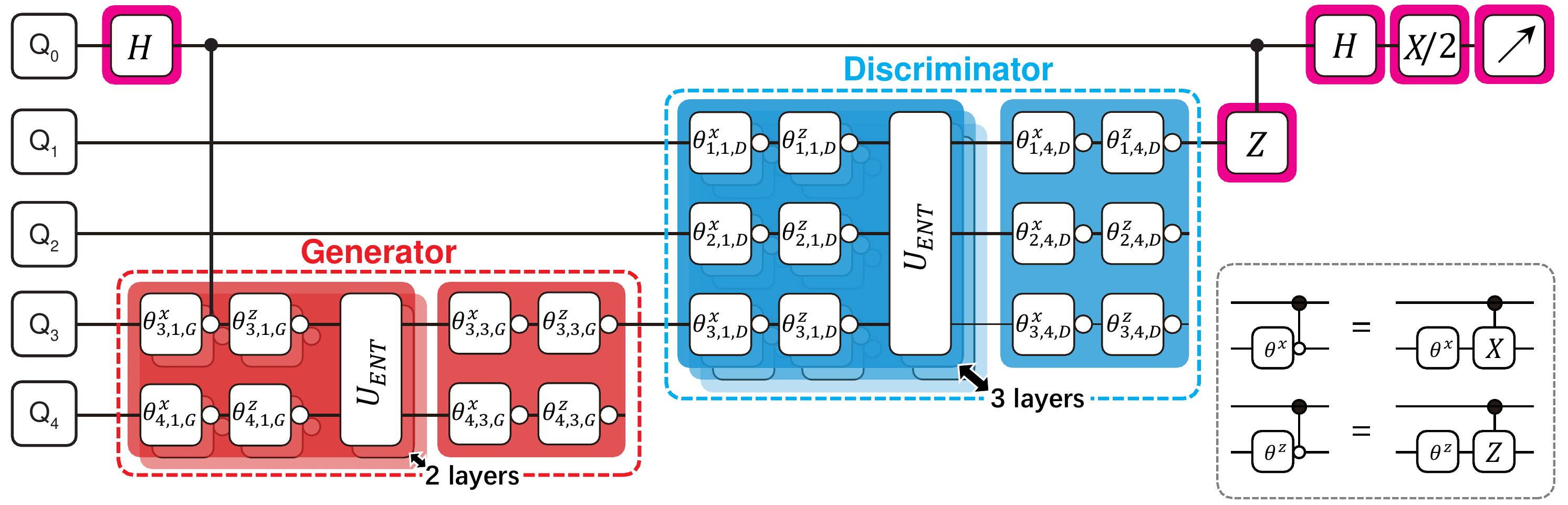}
    \caption{A concrete example of $\QTQGQD$ to learn the classical XOR gate. The generator and the discriminator are all parameterized quantum circuits, hardware-efficient Ans\"atze to be specific. 
    For $\theta^{x/z}_{i,j,m}$, the superscript ($x$ or $z$) refers to the axis in the Bloch sphere around which the state of $Q_i$ is rotated by the angle $\theta$ in the $i$-th layer.
    Image adapted from Ref.~\cite{huang2020realizing}.}
    \label{fig:QGAN-QT-QGQD}
\end{figure*}

\subsubsection{Applications}
QCBM and its variants have been applied in finance, such as learning empirical financial data distributions \cite{alcazar2020classical,kondratyev2021non}, generating synthetic financial data \cite{coyle2021quantum,kondratyev2021non}, and learning joint distributions \cite{coyle2021quantum,zhu2021generative}. Some recent works applied QCBM to real-life financial datasets and have shown better (or at least equivalent) performance than classical learning models like the RBMs \cite{kondratyev2021non,coyle2021quantum,alcazar2020classical,zhu2021generative}, especially when the scale of the problems increases.  \citet{zhu2021generative} reported an exponential advantage of QCBM with a SPSA optimizer over classical methods. Their approach utilized an Ansatz called `qupola' and learned the joint probability distribution of two random variables of the stock market by modeling their copulas using the IonQ trapped-ion platform with up to 8 qubits. It is also possible to utilize QCBMs to efficiently prepare quantum states of interested financial data \cite{pistoia2021quantum,stamatopoulos2020option} when certain conditions, e.g., the integrable property \cite{grover2002creating} are satisfied.

\subsection{Quantum generative adversarial network}
\label{sec:QGAN}

The concept of the quantum generative adversarial network (QGAN) was first conceived by \citet{lloyd2018quantum}. The underlying mechanism of QGAN accords with classical GANs introduced in Section \ref{subsec:classical_GAN}, where the generator and the discriminator are employed to carry out a two-player minimax game. QGAN can be used to estimate both discrete and continuous distributions. The key difference between QGAN and its classical counterparts is how the generator and discriminator are constructed. As depicted in Fig.~\ref{fig:GAN}, classical GANs generally employ deep neural networks to implement these two players, while QGANs may exploit QNNs. Celebrated by the strong power of quantum computers, such a construction rule may enable QGANs to gain certain computational advantages over classical GANs. In this regard, Ref.~\cite{lloyd2018quantum} discussed four types of QGANs, categorized by the learned tasks and the ways of implementing generators and discriminators. In what follows, we separately summarize these four types of QGANs and discuss their merits and limitations. 

The first type of QGAN, abbreviated as $\QTQGQD$, orients to the $\mathsf{Q}$uantum $\mathsf{T}$ask and consists of both $\mathsf{Q}$uantum $\mathsf{G}$enerator and $\mathsf{Q}$uantum $\mathsf{D}$iscriminator. In this scenario, the generator $G$ and the discriminator $D$ refer to QNNs $U(\vect{\theta_G})$ and $U(\vect{\theta_D})$, respectively. The aim of $G$ is to generate a quantum state $\rho = U(\vect{\theta_G}) \rho_0 U(\vect{\theta_G})^{\dagger}$, where $\rho_0$ is the initial input state of $G$, to approximate the target state $\sigma$. Conversely, the goal of $D$ is to distinguish the target state $\sigma$ from the generated state $\rho$. The objective function $\mathcal{C}_\text{QGAN}(\vect{\theta_G}, \vect{\theta_D})$ for $\QTQGQD$ \cite{benedetti2019quantum} can be defined as  
\begin{equation}\label{eqn:loss_QTQGQD}
	 \text{tr}(\Pi U(\vect{\theta_D}) \sigma U(\vect{\theta_D})^{\dagger}) p_t  - \text{tr}(\Pi U(\vect{\theta_D})\rho  U(\vect{\theta_D})^{\dagger}) p_g, 
\end{equation}
where $\Pi$ is the predefined measurement operator, and $p_t$ and $p_g$ refer to prior of feeding $\sigma$ and $\rho$ to $D$ with $p_t + p_g=1$. The optimization of QGAN follows $\min_{\vect{\theta_G}}\max_{\vect{\theta_D}}\mathcal{C}_\text{QGAN}(\vect{\theta_G}, \vect{\theta_D})$, which can be completed by using the methods introduced in Section \ref{subsec:vqa}. Specifically, QGANs are trained in a two-step strategy: the discriminator is updated first to obtain a better judgement while the generator fixed; then the generator is updated to be more deceptive while the discriminator is fixed. These two steps are conducted successively  until $\mathcal{C}_\text{QGAN}$ converges. Following optimization, the discriminator $D$ labels the input state as true if the posterior probability of $\sigma$ is greater than that of $\rho$ and vice versa. Fig.~\ref{fig:QGAN-QT-QGQD} shows how to use $\QTQGQD$ to learn the classical  XOR  gate. Since the generated data refers to the density matrix, which can be interpreted as a discrete probability distribution, it can be concluded that $\QTQGQD$ is designed for estimating discrete distributions.  

The theoretical foundation of $\QTQGQD$ is established in Ref.~\cite{lloyd2018quantum}. Due to Naimark's dilation theorem, the role of the discriminator $U(\vect{\theta_D})^{\dagger}\Pi U(\vect{\theta_D})$ corresponds to a positive operator-valued measurement (POVM) $\{\Pi_T, \Pi_F\}$ with $\Pi_T+\Pi_F=\mathbb{1}$.  As such, for a fixed $G$, the minimization of $\mathcal{C}_\text{QGAN}$ in Eqn.~(\ref{eqn:loss_QTQGQD}) with respect to $D$ can be achieved by the Helstrom measurement  \cite{helstrom1969quantum}.  The probability that the input data is sampled from the target distribution (or produced by the generator) refers to $p(T |\sigma) = \tr{(\Pi_T\sigma)}$ (or $p(T |\rho) = \tr{(\Pi_T\rho)}$). Moreover, the Nash equilibrium $\mathcal{C}_\text{QGAN}=0$ can only be achieved when the generated state exactly recovers the target state with $\rho=\sigma$. Moreover, as shown in Ref.~\cite{lloyd2018quantum}, if we can acquire the gradient information of $p(T |\sigma)$ or $p(T |\rho)$ through the  convex strategy space  at each iteration, the discriminator or generator can move directly towards the Nash equilibrium.

The second type of QGAN, abbreviated as $\QTCGQD$, targets the $\mathsf{Q}$uantum $\mathsf{T}$ask and consists of $\mathsf{C}$lassical $\mathsf{G}$enerator and $\mathsf{Q}$uantum $\mathsf{D}$iscriminator.  Considering that some distributions generated by quantum circuits (e.g., the output of IQP circuit \cite{bremner2016average} and boson sampling \cite{aaronson2011computational}) cannot be efficiently simulated by classical machines, classical $G$ may not fit the quantum target distribution precisely.  Consequently, following the theory of $\QTQGQD$, there exists a measurement for the discriminator to distinguish the target distribution from the generated distribution, and the Nash equilibrium can never be achieved.

The third type of QGAN, abbreviated as $\QTCGCD$, targets the $\mathsf{Q}$uantum $\mathsf{T}$ask and consists of both $\mathsf{C}$lassical $\mathsf{G}$enerator and $\mathsf{D}$iscriminator. Although the classical $G$ may not exactly recover the target quantum data, the classical $D$ may fail to detect the difference between the generated data from a classical generator and the complex quantum target data. To this end, the Nash equilibrium can still be achieved. Note that $\QTCGCD$ is closely associated with the topic of using deep neural networks to study quantum physics problems, including quantum state tomography and the entanglement identification \cite{carleo2019machine,dunjko2018machine}.

The last type of QGAN discussed in Ref.~\cite{lloyd2018quantum}, abbreviated as $\CTQGCD$, targets the $\mathsf{C}$lassical $\mathsf{T}$ask and consists of $\mathsf{Q}$uantum $\mathsf{G}$enerator and $\mathsf{C}$lassical $\mathsf{D}$iscriminator. $\CTQGCD$ is designed for estimating continuous distributions, where the generated data is described by the expectation values when interacting the quantum state with different measurement operators.  As claimed by the authors, $\CTQGCD$ may possess computational advantages over classical GANs, warranted by the ability of quantum processors to represent $N$-dimensional vectors using $\log N$ qubits. In this way, advanced quantum algorithms can apply linear algebra on those vectors in time $O(poly(log N))$ to optimize both the generator and discriminators. By contrast, classical GANs generally request $O(poly(N))$ runtime complexity in optimization.  

\begin{table}[H]
	\centering
	\caption{Variants of QGANs and their abbreviations. For example, $\QTCGQD$ stands for quantum target data, classical generator, and quantum discriminator.
	}
	\label{table:qgan}    
    \begin{tabular}{c|cc}
        \toprule
        \diagbox{Generator}{Discriminator} & Quantum & Classical \\
        \hline
        Quantum (CT) & $\CTQGQD$  & $\CTQGCD$   \\
        Quantum (QT) & $\QTQGQD$ & $\QTQGCD$ \\
        Classical (CT) & $\CTCGQD$ & $\CTCGCD$ \\
        Classical (CT) & $\QTCGQD$ & $\QTCGCD$  \\
        \bottomrule
    \end{tabular}
\end{table}

\subsubsection{Variants}

The development of powerful QGANs has attracted great attention in the past three years. To elucidate, here we follow the method used in Ref.~\cite{lloyd2018quantum} to categorize QGANs. Specifically, according to whether quantum resources are involved in the learning task or the construction of learning models, previous literature related to QGANs can be cast into eight groups, as summarized in Table \ref{table:qgan}. Considering that the main focus of this 
work is quantum learning models, we exclude the discussion about $\CTCGCD$ and $\QTCGCD$.

\underline{\textit{$\QTQGQD$}.}
The potential of $\QTQGQD$ has been extensively explored from both algorithmic and experimental aspects. In particular, \citet{dallaire2018quantum} proposed the quantum extension of conditional GANs \cite{DBLP:journals/corr/MirzaO14}, dubbed QuGAN, and claimed that the proposed model has a higher representation power than their classical counterparts. Potential applications of QuGAN include performing quantum chemistry calculations on quantum computers and compressing time evolution gate sequences.     \citet{benedetti2019adversarial} utilized QGAN to estimate an unknown quantum pure state on near-term quantum computers. They suggested using resilient back-propagation as the optimizer and employing the entanglement entropy as the stopping criterion. Relating to the same task, \citet{hu2019quantum} experimentally demonstrated the first proof-of-principle QGAN on superconducting quantum devices to approximate a single-qubit state.  Then, \citet{huang2020realizing} implemented QGAN on a 5-qubit superconducting quantum device to learn multi-qubit quantum states and the classical XOR gate.  \citet{yin2022efficient} proposed a quantum adversarial solver to address quantum information processing tasks. By restricting the expressive power of the quantum generator, their proposal can efficiently detect the bipartite entanglement of quantum states. Experimental results on the linear optical network indicated the effectiveness of the proposal and exhibited potential quantum advantages. On par with designing different implementations of QGANs, some studies advised using different cost functions during training to attain robust performance.  \citet{chakrabarti2019quantum} studied the smoothness of the original cost function and proposed the Wasserstein semi-metric for quantum data, based on which they constructed the quantum WGAN model.  \citet{stein2020qugan} proposed a quantum-based cost function, called Quantum State Fidelity, to reduce the number of parameters. 

\begin{figure}[htbp]
    \centering
    \includegraphics[width=0.48\textwidth]{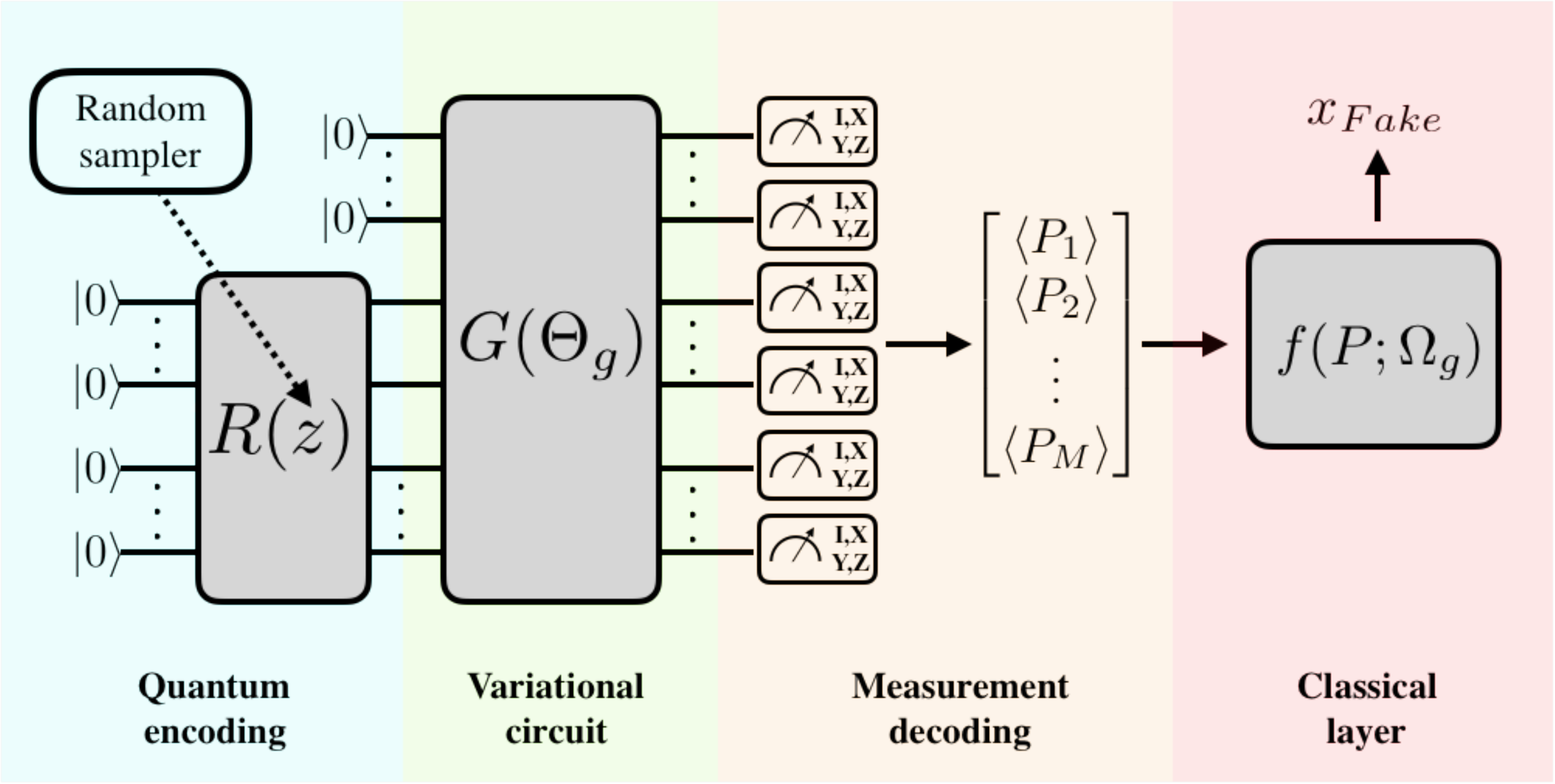}
    \caption{A quantum generator for continuous distribution.
    The random variable $z$ is mapped to a quantum state by a data encoder $R$. 
    The output of $G$ in a quantum state is measured multiple times for an approximate distribution representation. And before input to the discriminator, a classical layer produces $x_{Fake}$ as generated data in the continuous space.
    Figure adapted from Ref.~\cite{romero2021variational}.}
    \label{fig:QGAN-continuous}
\end{figure}

\underline{\textit{$\QTQGCD$ \& $\CTQGCD$.}} A large body of works falls into the regime of $\QTQGCD$ and $\CTQGCD$. In particular, \citet{zoufal2019quantum} proposed a type of QGAN that aims to efficiently learn and load random distributions. Experiments conducted on the IBMQ cloud validated the effectiveness of their proposal. \citet{zeng2019learning} presented a QGAN model, as an adversarial training scheme for QCBM. Compared with the original QCBM, their proposal quadratically reduces the sample complexity of estimating the cost function. The proposed model can be applied to infer the unobserved data conditioned on partial observations such as image in-painting.   \citet{nakaji2021quantum} proposed the quantum semi-supervised GAN (QSGAN). As with the classical semi-supervised GANs that train the discriminator as a good classifier \cite{odena2016semi}, the main focus of QSGAN is training a good classical discriminator rather than the quantum generator. The simulation results showed that QSGAN can achieve comparable performance to classical semi-supervised GANs by using a fewer number of parameters.  \citet{situ2020quantum} designed a type of QGAN and demonstrated its potential to avoid the vanishing gradients problem encountered in classical GANs when generating discrete data distributions. \citet{romero2021variational} proposed a variational quantum generator, as shown in Fig.~\ref{fig:QGAN-continuous}, to model continuous classical probability distributions.  \citet{li2021quantum} demonstrated how to use a quantum generator to significantly reduce the number of required parameters compared to its classical counterpart in the regime of drug discovery.

\underline{\textit{$\QTCGQD$ \& $\CTCGQD$ \& Others.}}
\citet{niu2021entangling} proposed an entangling QGAN model to overcome the issue of non-convexity and mode collapse. Despite adopting a quantum generator or a quantum discriminator or both, there are also works that add an auxiliary quantum neural network to assist the training of GANs. Associative adversarial networks (AAN) have been proposed to circumvent the imbalance in the training rates of generators and discriminators. Rather than using the uniform or Gaussian distribution in standard GANs, the prior distribution is gained from an additional RBM which is trained simultaneously with generators and discriminators. This RBM takes an intermediate layer of the discriminator as input. The authors claimed that sampling the distribution of the RBM is more meaningful than random noise. Refs.~\cite{wilson2021quantum,anschuetz2019near} extended classical AAN to quantum-classical hybrid architecture by transforming the associated RBM into QBM while leaving generators and discriminators classical. Ref.~\cite{zoufal2019quantum} explored how the varied optimizers, e.g., AMSGrad, another adaptive-learning-rate, gradient-based optimizer, affect the performance of QGANs. Ref.~\cite{kiani2021quantum} investigated different definitions of quantum Wasserstein distance and provided the implementation of quantum Wasserstein GAN.

Note that quantum generative adversarial networks discussed in this paper are different from another field called quantum adversarial learning which studies the effect of adversarial examples on a vulnerable model. Adversarial attack works by introducing certain perturbances, imperceptible to human eyes, to a normal sample, leading to a surprising result. Ref.~\cite{lu2020quantum} reasons about some adversarial scenarios in the context of quantum machine learning.

\subsubsection{Applications}

The potential applications of QGANs include quantum state tomography \cite{d2003quantum}, finance, image generation, and drug discovery. In particular, \citet{benedetti2019adversarial} applied the proposed $\QTQGQD$ to approximate an unknown quantum pure state. In the domain of finance, \citet{zoufal2019quantum} applied QGANs to help financial derivative pricing and \citet{assouel2021quantum} applied QGAN models for volatility modeling. \citet{huang2021experimental} proposed patch-QGAN and batch-QGAN and experimentally demonstrated that these two strategies can generate images of real-world hand-written digits on super-conducting quantum machines. Last, the drug discovery science community has attempted to employ QGANs for drug discovery \cite{li2021quantum,li2021drug}, which employ the patch strategy.

\subsection{Quantum Boltzmann machine}
\label{QBM}
\begin{figure}[htbp]
    \centering
    \includegraphics[width=0.3\textwidth]{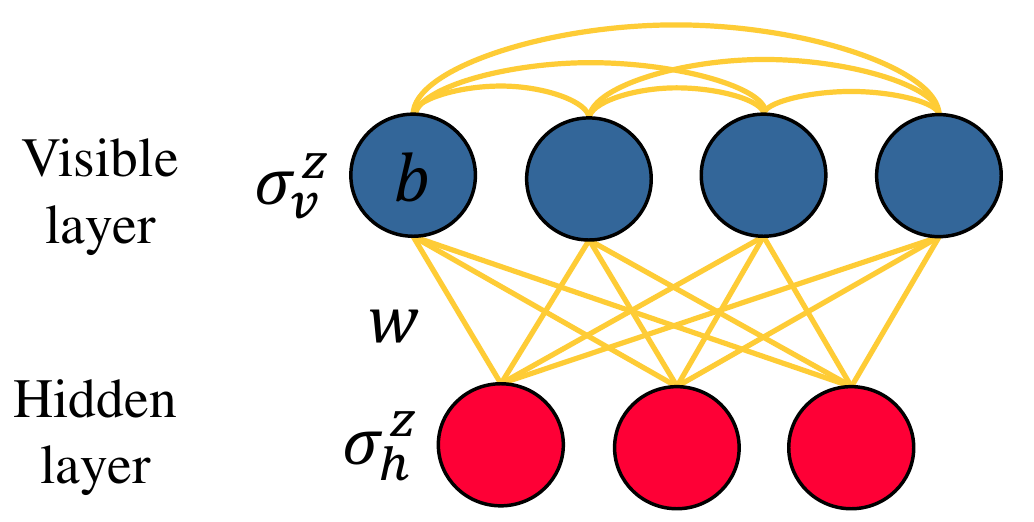}
    \caption{A semi-RQBM model with variables of visible layer ($\vect{\sigma^z_v}$) and hidden layer ($\vect{\sigma^z_h}$) respectively.
    Compared to the quantum counterpart of traditional RBM (as shown in Fig.\ref{fig:BM}, right panel), where all internal connections within visible and hidden units are disallowed, semi-RQBM is more suitable for further analysis. A transverse filed $\Gamma$ is also present in every unit to represent the noncommutative part in the Hamiltonian of the system.}
    \label{fig:QBM}
\end{figure}

The quantum Boltzmann machine (QBM) is first proposed by \citet{amin2018quantum}. As the quantum generalization of the classical Boltzmann machine (BM) introduced in Section~\ref{sec:clc_bm}, it exploits quantum devices to prepare the Boltzmann distribution estimating the target discrete distributions. The main difference between QBM and classical BMs is that the units in Fig.\ref{fig:BM} are replaced by qubits and naturally the energy term appeared in Eqn. \eqref{bm-prob} is replaced by a Hamiltonian. For a transverse-field Ising model, the Hamiltonian takes the form $H (\vect\theta)= - \sum_{i, j} w_{i j} \sigma_{i}^{z} \sigma_{j}^{z} - \sum_{i} b_{i} \sigma_{i}^{z} -\sum_{i} \Gamma_{i} \sigma_{i}^{x}$, where $\vect\theta=\{\vect{w}, \vect{b}, \vect{\Gamma}\}$ are the trainable parameters.  The implementation of a semi-restrict QBM (semi-RQBM) is depicted in Fig.~\ref{fig:QBM}, where the internal connection between hidden units are disallowed. For a general QBM, the probability of observing the state $|\vect{v}\rangle$ is $p_{\vect\theta}(\vect{v})= \tr (\Lambda_{\vect{v}} \rho)=Z^{-1}\tr (\Lambda_{\vect{v}} e^{-H (\vect\theta)})$, where $Z=\tr (e^{-H (\vect\theta)})$ is the partition function, $\rho = Z^{-1}e^{-H (\vect\theta)}$ is the density matrix, and the diagonal matrix $\Lambda_{\vect{v}}=|\vect{v}\rangle\langle\vect{v}| \otimes \mathbb{1}_{\vect{h}}$ is a standard projective measurement.  To approximate the target distribution $q$, QBM updates its trainable parameters $\vect{\theta}$ to minimize the negative log-likelihood, i.e., 
     $\mathcal{C}_{\text{QBM}} = -\sum_{\vect{v}} q(\vect{v}) \log {p(\vect{v})}$. The gradients of $\mathcal{C}_{\text{QBM}}$ take the form 
\begin{equation}
\label{qbm_differential}
    \partial_{\vect\theta} \mathcal{C}_{\text{QBM}}=\sum_{\vect{v}} q(\vect{v})\left(\frac{\tr (\Lambda_{\vect{v}} \partial_{\vect\theta} e^{-H})}{\tr (\Lambda_{\vect{v}} e^{-H})}-\frac{\tr (\partial_{\vect\theta} e^{-H})}{\tr (e^{-H})}\right),
\end{equation}
where the first term is called the positive phase and the second term (\emph{negative phase}) refers to the Boltzmann average and is equivalent to $\left\langle\partial_{\vect\theta} H\right\rangle =\tr (\rho \partial_{\vect\theta} H)$. Note that since the exact sampling of $\rho$ is NP-hard \cite{barahona1982computational}, the calculation of the negative phase becomes intractable for large-scale problems. To this end, instead of exact sampling, quantum analogues of the classical constructive divergence method \cite{hinton2002training} have been proposed \cite{poulin2009sampling,yung2012quantum}.  By leveraging the Golden-Thompson inequality \cite{golden1965lower,thompson1965inequality},
\citet{amin2018quantum} argued that the training cost of QBM can be replaced by its upper bound $\tilde{\mathcal{C}}_{\text{QBM}}$, i.e., ${\mathcal{C}}_{\text{QBM}} \leq \tilde{{\mathcal{C}}}_{\text{QBM}} \equiv -\sum_{\vect{v}} q(\vect{v})\log [\tr ( e^{-H_{\vect{v}}})/\tr ( e^{-H} )]$,
where $H_{\vect{v}}=\langle\vect{v}|H| \vect{v}\rangle$ is the clamped Hamiltonian. In this way, the trainable parameters $\{\vect{b}$, $\vect{w}\}$ can be updated efficiently. However, this method does not allow the training of the transverse filed $\vect{\Gamma}$, which has to be preset as a hyper-parameter in this model.

\citet{amin2018quantum} further applied two types of QBMs, i.e., the fully visible QBM and the semi-restricted QBM (semi-RQBM), to simulate a Bernoulli mixture distribution. Experiments have shown that QBMs achieve better trainability than classical ones. A supervised generative task was also performed on a fully visible model, which takes an 8-qubit input and provides a 3-qubit output, where the input $\vect{x}$ and the output $\vect{y}$ are treated jointly as the visible pairs $\vect{v}=(\vect{x},\vect{y})$.  The numerical result indicated that QBM trained with a joint distribution $p(\vect{x}, \vect{y})$ obtained a much lower cost over classical BMs. The authors also discussed the potential of employing a quantum annealer as the processor to implement QBM. To this end, \citet{benedetti2017quantum} experimentally confirmed the feasibility of this proposal by implementing BM models on the D-Wave 2X quantum annealer with up to 940 qubits to accomplish unsupervised generative learning tasks.

The most problematic part of QBM models is the sampling of the positive and negative phases respectively. Regarding the positive phase, it could be numerically estimated using exact diagonalization or quantum Monte Carlo (QMC) techniques, or more efficiently, be approximated via the Golden-Thompson inequality \cite{anschuetz2019realizing,amin2018quantum,xiao2020quantum}. However, it prevents the training of the transverse field and hence that has to be treated as a hyper-parameter. Though employing POVM  \cite{kieferova2016tomography} mitigates this issue, it requires prior information on the data distribution and is generally difficult. 
There might be other approximation methods that yield better performance like quantum mean-field methods \cite{xiao2020quantum}. 
The negative phase sampling is a computationally NP-hard problem where the exact diagonalization needs exponential complexity. 
Multiple works have been proposed to accelerate the sampling of the Gibbs state. The general approach to approximate thermal state sampling is by applying QMC techniques, including simulated quantum annealing (SQA) \cite{crawford2016reinforcement} and the state-of-art CT-QCMC \cite{xiao2020quantum}. 
With no known theoretical bound complexity, heuristic algorithms for thermodynamics or other physics systems can hopefully find more efficient solutions. For instance, Ref.~\cite{anschuetz2019realizing} adopted a combination system based on the eigenstate thermalization hypothesis, while it is not applicable for general interaction systems \cite{rigol2008thermalization}. Ref.~\cite{zoufal2021variational} proposed a method using VarQITE and experimentally implemented QNN on quantum hardware, which provides a framework of quantum circuits with complex structures.

\subsubsection{Variants}

The variants of QBMs can be divided into three classes, depending on the training method. The first class is bound-based training in which the original cost function is replaced by its upper bound \cite{amin2018quantum,PhysRevA.96.062327,anschuetz2019realizing,khoshaman2018quantum}. The second class is the state-based training by minimizing the relative entropy \cite{kappen2020learning,PhysRevA.96.062327,wiebe2019generative,wiebe2019quantum}. The last class is the circuit-based training \cite{zoufal2021variational}, which utilizes variational quantum imaginary time evolution (VarQITE) methods \cite{magnus1954exponential,mcardle2019variational,yuan2019theory}.
 
\underline{\textit{Bound-based training}}. The key concept of the bound-based training is using Golden-Thompson inequality to impose an upper bound on the cost function. This allows an efficient calculation of the gradient information. \citet{PhysRevA.96.062327} proposed a POVM-based training scheme which is a generalization of Ref.~\cite{amin2018quantum}, where the measurements $\Lambda_{\vect{v}}$ in  Eqn.~\eqref{qbm_differential} are generalized to POVM. As such, the training in the transverse field can proceed. To explore potential quantum advantages, Ref.~\cite{PhysRevA.96.062327} applied QBMs to solve Hamiltonians with the fermionic form since the sign problem \cite{hangleiter2020easing,li2015solving} prevents the efficient implementation of QMC techniques. Simulation results have indicated that QBMs can provide a significantly enhanced learning performance over classical BMs. 

\citet{puvskarov2018machine} argued that the generalized Gibbs ensembles \cite{rigol2007relaxation,barthel2008dephasing} could be leveraged as a basis for QBMs. In doing so, the gradient information can be effectively obtained due to the commutativity of the conserved charge in the integrable Hamiltonian by learning the optimal effective temperatures. Simulation results showed that their proposal could achieve a reasonably low error rate with much fewer parameters than RBM. \citet{anschuetz2019realizing} proposed an efficient method to approximately sample the  thermal states of the QBM negative phase, which is supported by the eigenstate thermalization hypothesis \cite{srednicki1994chaos,deutsch1991quantum}. The numerical simulation results illustrated that their proposal has a comparable performance with the exact QBM and is superior to the classical RBM. Meanwhile, it is also robust against noise.  \citet{xiao2020quantum} proposed a general training scheme for various QBMs, which includes  general transverse-field Ising models and the stochastic $\kappa$-local ($\kappa \geq 2 $) Hamiltonians. The authors also verified the efficiency of their proposal through simulations on small-scale multi-mode Bernoulli distribution \cite{anschuetz2019realizing,juan2004bernoulli} and large-scale MNIST dataset.

\underline{\textit{State based training}}. \citet{PhysRevA.96.062327} first illustrated a state-based training scheme for QBM  without hidden units. The model is trained to minimize the relative entropy between a known quantum state $\rho$ prepared by an oracle and the generated observable state with density matrix $\sigma$, i.e., $S\left(\rho|\sigma\right)=\tr (\rho \log \rho)-\tr \left(\rho \log \sigma\right)$ \cite{carlen2010trace}. This protocol provides an alternative technique to quantum state tomography (QST) and can further generate copies of the learned quantum distributions. Based on the fully visible QBM model, the tomographic reconstructions of Haar-random pure states and mixed two-qubit states show a high level of accuracy using the state-based training. \citet{wiebe2019generative} provided two training methods for QBM to minimize the relative entropy with the presence of hidden units. The visible part of the generated observable state $\sigma$ takes the form $\sigma_{v}(H)={\tr _{h}( e^{-H}})/{\tr (e^{-H})}$. The first approach is to propose a variational upper bound of the cost function, which is based on a constrain-form Hamiltonian with commutative hidden units. To obtain the gradient of this upper bound, the Gibbs state preparation, Hadamard test, and amplitude estimation subroutines are needed. The second approach is adaptive to general forms of Hamiltonians and is more up-to-date. A truncation on the term $\text {log} \sigma_{v}$ \cite{van2020quantum} is taken 
and then Fourier series approximation is employed with $\sigma_{v}\approx \sum_{m} c_{m} \exp \left(i m \pi \sigma_{v}\right)$. In conjunction with the divided difference method, the gradients of the cost function can be efficiently estimated and the evaluation of which could be realized through the Hadamard test and sample-based Hamiltonian simulation \cite{lloyd2014quantum,kimmel2017hamiltonian} subroutines that are available on quantum devices.

\underline{\textit{Quantum circuit based training}}. \citet{zoufal2021variational} proposed a variational QBM. Different from previous literature, the specified Hamiltonian takes a distinct form, i.e.,  $H_{\vect\theta}=\sum_{i=0}^{p-1} \vect\theta_{i} h_{i} \equiv \sum_{i=0}^{p-1} \vect\theta_{i} \bigotimes_{j=0}^{n-1} \sigma_{j, i}$, where $\vect\theta \in \mathbb{R}^p$ and the Pauli operator $\sigma_{j, i} \in\{I, X, Y, Z\}$ acts on the $j$-th qubit. Utilizing the VarQITE method, the approximated Gibbs states are prepared by quantum circuits, and the analytic gradients are computed using automatic differentiation method. The effectiveness of the VarQITE method is confirmed on quantum hardware by examining the closeness of the prepared Gibbs states and the target ones, and the experimental results on generative and discriminative tasks also demonstrated the feasibility and the noise-resilient property of variational QBMs.

\subsubsection{Applications}
\citet{10.5555/3370185.3370188} employed RBM, deep Boltzmann machine (DBM), and QBM models of clamped Hamiltonians to accomplish RL tasks of Maze Traversal problems \cite{sutton1990integrated}, which are Markov decision processes and are widely adopted to benchmark RL algorithms. As it is argued that the simulated quantum annealing (SQA) \cite{crosson2016simulated,heim2015quantum} could simulate the quantum annealing process due to its  behavioral similarity with quantum tunneling and entanglement \cite{albash2015reexamining,isakov2016understanding,shin2014quantum}, it is utilized to generate samples approximating Boltzmann distributions. In the numerical simulations, the SQA method is applied to both the classical Hamiltonian of DBM (by setting the strength of the transverse field in the freeze-out point \cite{amin2015searching} $\Gamma_f \approx 0$ ) and the quantum Hamiltonian of QBM (by setting $\Gamma_f > 0$ and employing an effective Hamiltonian one dimensional higher). The results showed that the DBM outperforms RBM as the number of training samples increases, whereas the QBM has the best performance. \citet{wiebe2019quantum} exploited a Fock-space representation to encode linguistic structures in vector spaces.  In this way, the authors illustrated how QBM could be applied to solve language processing tasks.

\subsection{Quantum autoencoder}
The concept of quantum autoencoder (QAE) was first proposed by two pioneering works \cite{wan2017quantum,romero2017quantum}. QAE inherited the spirit of classical autoencoders as described in Section \ref{sec:clc_ae}. Namely, the central concept of QAE is compressing the input quantum state into a low-dimensional latent space such that the input can be recovered from this dimension-reduced representation.  Two core components of QAE are the \textit{encoder} and the \textit{decoder}, which are implemented by QNNs  \cite{romero2017quantum}. The optimization of QAE amounts to minimizing the reconstruction error between the input quantum state and its reconstruction. Since QAE does not encounter the read-in and read-out bottleneck, it possesses potential computational advantages in the task of quantum information processing and quantum many-body physics.

\begin{figure}[ht]
    \centering
    \includegraphics[width=0.45\textwidth]{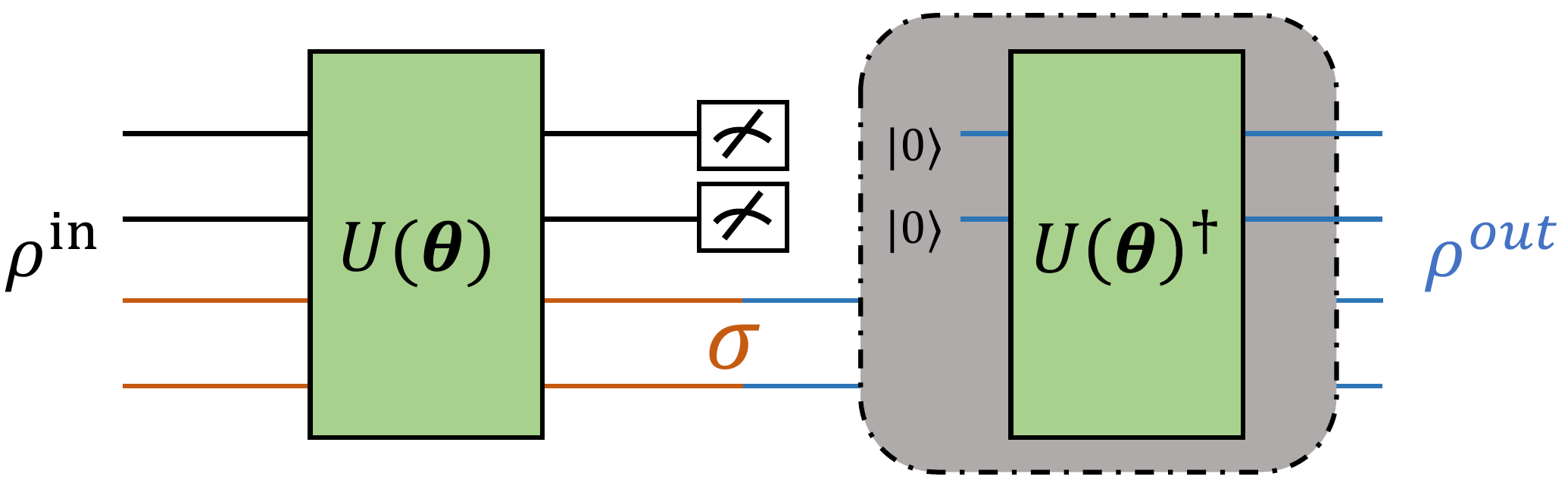}
    \caption{The structure of a QAE. The red lines and the black lines represent the qubits of subsystem $A$ and subsystem $B$ respectively. 
    The trash state of $B$ is measured before a reference state takes the place of the trash state. The compressed state of subsystem $A$ is denoted by $\sigma$.
    Note that the reference state is not necessarily restricted to be in state $|0\rangle^{\otimes n}$.  
    Image adapted from Ref.~\cite{du2021exploring}.
    }
    \label{fig:QAE}
\end{figure}

The structure of QAE \cite{romero2017quantum} is depicted in Fig.~\ref{fig:QAE}. In particular, an input state $\rho^{in}=|\psi_i\rangle \langle\psi_i|$ sampled from an ensemble of $(n_A+n_B)$-qubit states $\{p_i, |\psi_i\rangle\}$ is fed into the encoder $U({\vect\theta})$, where $A$ and $B$ are two subsystems with $n_A$ and $n_B$ qubits respectively. Applying the partial measurement to the subsystem $B$, the compressed state is $\sigma=\tr_B(U({\vect\theta})\rho^{in}U({\vect\theta})^{\dagger})$. This completes the encoding process.  Then, by interacting $\sigma$ with another subsystem $B'$ followed by operating the decoder $U^\dagger(\vect\theta)$, the recovered state takes the form  $\rho^{out}= U^{\dagger}(\vect\theta)(\sigma \otimes |0\rangle\langle 0|^{\otimes n_{B'}}) U(\vect\theta)$. The cost function of QAE is defined as the weighted average infidelity between the input ${\rho^{in}}$ and the output $\rho^{out}$, i.e., 
\begin{equation} \label{c1}
    \mathcal{C}_{\text{QAE}}(\vect\theta)=1- \sum_{i} p_{i} \cdot F(\rho^{in}, \rho_{i}^{out}),
\end{equation}
where $F\left(\rho^{in}, \rho^{out}_{i}\right)=\tr(\rho^{in}_i\rho^{out}_i)$ is the expected fidelity \cite{wilde2013quantum}. When $   \mathcal{C}_{\text{QAE}}=0$, the state ${\rho^{out}}$ exactly recovers the input $\rho^{in}$. From the view of generative learning, QAE is applied to estimate the discrete distribution (formalized by the density matrix $\rho^{in}$).

The authors applied the proposed QAE to tackle three learning tasks, i.e.,  reconstructing the ground states of molecular hydrogen, Hubbard models, and $H_4$ molecular. The implementation of QAE is achieved by adopting programmable quantum circuits \cite{daskin2012universal,10.5555/2011717.2011721}. The L-BFGS-B and basin-hopping \cite{wales1997global} optimizers are employed to minimize the cost function. The simulation results indicated the efficiency of compressing quantum states. 

\subsubsection{Variants}
The variants of QAEs can be categorized by their applications. In the following sub sections, we separately introduce QAEs designed for quantum state compression, reconstruction, preparation, and denoising.

\underline{\textit{Quantum states compression}}. An efficient method of state compression can dramatically save the required quantum memory and thus has a practical value in quantum chemistry simulation \cite{peruzzo2014variational,o2016scalable} as discussed in Refs.~\cite{romero2017quantum,bravo2021quantum}. Due to this reason, there is a growing interest in designing powerful QAEs to complete state compression tasks. In particular,  \citet{bravo2021quantum} proposed an enhanced feature quantum autoencoder (EF-QAE) with local cost functions to avoid the trainability issues \cite{cerezo2021cost,mcclean2018barren}  incurred by global cost functions. The authors adopted an encoding strategy that encodes the feature vector of the input data into every single-qubit rotation gate implemented in QNNs \cite{perez2020data}. The simulation results of compressing the ground states of transverse-field Ising (TFI) Hamiltonian \cite{bravo2021quantum} and classical handwritten digits showed that EF-QAE could reach a higher fidelity than standard QAEs. Besides,  \citet{srikumar2021clustering} proposed a hybrid QAE with a  QFFNN-ANN-QFFNN  structure, where the  quantum forward neural network (QFFNN)  \cite{beer2020training} is implemented by coupling QNNs and measurements. \citet{huang2020realization} proved that an optimal compression can never occur if the rank of the input states is greater than the dimension of the compressed latent space. \citet{ma2020compression} proposed a closed-loop learning control strategy to optimize QAE, where the trainable parameters refer to the control field of the encoding unitary. Training QAE with four different optimizing control methods \cite{khaneja2005optimal,judson1992teaching,dong2019learning,salimans2017evolution} are simulated and compared.

Several studies have explored the possibility of using QAE to complete state compression on photonic systems \cite{pepper2019experimental,huang2020realization,ma2020compression,steinbrecher2019quantum}. Specifically, \citet{steinbrecher2019quantum} integrated a quantum optical neural network (QONN) into QAE. The proposed optical QAE can compress quantum optical states by imposing layers of linear optical unitary and single-site nonlinearities with interaction strength.  \citet{pepper2019experimental} performed experiments with QAE based on the control of a photonic device. The results demonstrated that  QAE could achieve low error rates when compressing qutrits to qubits. Moreover, the proposed QAE is effective and robust during optimization under channel drift. 

\underline{\textit{Quantum states reconstruction}}. Analogous to classical autoencoders, QAEs have been used to reconstruct quantum states, which is important in quantum communications \cite{wan2017quantum,zhang2021chip}. Particularly, \citet{wan2017quantum} proposed a theoretical scheme to realize the quantum-neural-based QAEs and suggested a quantum communication protocol. Their proposal has the same underlying structure as the autoencoder of FFNN.  Similar to Eqn.~\eqref{c1}, the cost function quantifies the differences between the input and output states through Pauli gate measurements.  The authors also discussed the physical feasibility of QAEs via the proposed quantum optical neuron module. Recently, \citet{cao2021noise} have proposed a noise-assisted quantum autoencoder (N-QAE) to recover general quantum states, including high rank and mixed states. Since the perfect compression cannot be achieved when the rank of $\rho^{out}_{i}$ is larger than $n_A$ \cite{ma2020compression}, N-QAE employs devised amplitude-damping channels to prepare mixed reference states, which re-utilize the measurement information of the trash state. By doing this, the  N-QAE can increase the rank of the output state. The authors also designed an adiabatic model of quantum autoencoder (A-QAE) to handle large-scale problems without the vanishing gradient issue. Numerical experiments on reconstructing the thermal states of the TFI model and Werner states have verified the feasibility of A-QAE. \citet{zhang2021chip} first illustrated a QAE implementation on a silicon phonic chip, which facilitates the execution of a quantum-autoencoder-facilitated teleportation (QAFT) protocol. With QAFT, some high dimensional quantum states (e.g. qutrit) can be compressed into qubits on the transmitter chip and teleported to the receiver chip where the qutrit can be reconstructed with considerable fidelity.

\underline{\textit{Quantum states preparation}}. Another application of QAE is preparing certain quantum states that cannot be efficiently prepared by classical machines. \citet{du2021exploring} proved that the spectral information of the compressed state contains the spectral information of the input state. Supported by this property, the authors devised a method of using QAE to prepare the quantum Gibbs state \cite{brandao2017quantum,poulin2009sampling}  and estimate the quantum Fisher information. \citet{lamata2018quantum} established the connection between the approximate quantum adders \cite{li2017approximate} and QAEs. As the former enables encoding from a high  dimensional tensor product  state to one qubit state, it can be viewed as a trivial way to realize  QAE. To this end, a gate-limited quantum adder is proposed, where the internal structures of the encoder and decoder are optimized by the GA method. Subsequently, \citet{ding2019experimental} experimentally implemented the gate-limited quantum adders on a Rigetti cloud quantum computer with 2-qubit input states.  

\underline{\textit{Quantum states denoising}}.   \citet{bondarenko2020quantum} proposed a QAE based on  QFFNN  following the approach developed in Ref.~\cite{beer2020training}. The authors examined different topologies of QFFNNs via numerical simulations and showed that QAEs can denoise the GHZ states under bit-flip error and random unitary error. In the same vein, \citet{achache2020denoising} also exploited  QFFNN-based QAEs  to denoise quantum states.  \citet{zhang2021generic} proposed a detection-based QAE to conduct error mitigation. Their proposal does not require extra qubits and is resource-friendly to NISQ devices.  

\textit{\underline{Others}.} Besides the above applications, QAEs have been exploited to achieve classical tasks such as reconstructing handwritten digits \cite{bravo2021quantum,khoshaman2018quantum}, boosting high-dimensional search \cite{gao2020high}, and accomplishing   clustering \cite{srikumar2021clustering}. Furthermore, recent advances in photonic techniques \cite{chung2017monolithically,zhu2018scalable} provide an opportunity to use QAEs \cite{pepper2019experimental,steinbrecher2019quantum,huang2020realization,ma2020compression} to complete teleportation tasks with potential quantum advantages.  Further performance improvement can be deliberated in several aspects: taking hybrid schemes \cite{khoshaman2018quantum,gao2020high,srikumar2021clustering}, leveraging noise and quantum annealers \cite{cao2021noise}, harness the encoding strategy \cite{bravo2021quantum}, and taking different structures of encoder-decoder, e.g. QNNs and controlled unitary field \cite{ma2020compression}. The Ansatz may be delicately designed when utilizing PQCs, while the barren plateaus or narrow gorges \cite{cerezo2021cost,arrasmith2021equivalence} problem still exist in deep PQCs. Besides the exploratory Ansatz used in Refs.~\cite{romero2017quantum,zhang2021generic,bravo2021quantum,cao2021noise}, the hardware efficient Ansatz has also been tested \cite{du2021exploring,srikumar2021clustering}, and the GA algorithm could assist Ansatz designing \cite{lamata2018quantum,ding2019experimental}. When using QFFNN (QONN) circuits, improvement could be made concerning their topologies, circuit depth and training schemes. And being similar with the classical NNs, the issue of local minima may still exist when a lower depth of circuit is applied \cite{steinbrecher2019quantum}.

\citet{khoshaman2018quantum} proposed a quantum variational autoencoder (QVAE), by appending a QBM sub-model to represent the discrete latent space of the classical variational autoencoder.  With the QVAE model, \citet{gao2020high} explored an early applications in the earth science field to search high-dimensional similarity from satellite datasets. Implemented on   D-waves, the QVAE model attains   speedups and requires less storage space than classical methods.

\section{Conclusions and discussions}
\label{sec:conclusions}
In this survey, we provide a comprehensive review of numerous QGLMs and elaborate on their potential to benefit both  conventional machine learning tasks and  quantum physics from the perspective of computational efficiency and model expressivity.  Despite the astonishing progress, quantum generative learning is still in its infancy, and many critical issues still remain unexplored. In the remainder of this section, we discuss these unsolved issues and present insights into the future prospects of quantum generative learning.

\subsection{Challenges}

\textit{\underline{Read-in \& Read-out bottleneck}.} A common caveat of QGLMs is the read-in and read-out bottleneck presented in estimating  high-dimensional continuous distribution tasks \cite{aaronson2015read}. Particularly, both QAE and QGAN (i.e., $\QTCGQD$ and $\CTCGQD$) encounter the read-in bottleneck when the training data is classical and sampled from a continuous  distribution, and the amplitude encoding is specified. That is, how to efficiently load the example $\vect x\in \mathbb{R}^{2^N}$ into an $N$-qubit state $\ket{\vect x}=\frac{1}{\|\vect x\|_2}\sum_{i=1}^{2^N} \vect{x}_i \ket{i}$ remains unknown. The read-out bottleneck concerns how to extract quantum information into classical registers. In QGAN, when $\CTQGCD$ is applied to estimate the continuous distribution, it suffers from the read-out bottleneck. For example, in the task of image generation, the features of the generated image are stored in the probability amplitudes of quantum states. To extract these features into the classical world, QST is required, where the computational complexity scales with the feature dimension. As a result, current QGLMs are inferior at processing high-dimensional continuous distribution tasks. The design of efficient read-in and read-out strategies can dramatically improve the capabilities of QGLMs in the regime of many conventional machine learning tasks.

\textit{\underline{Learning machines}.}
 Another key challenge in the regime of QGLMs is exploring their learnability from the perspective of generalization, expressivity, and trainability. A deep theoretical understanding about these topics cannot only benefit the evaluation for the performance of different QGLMs but also provides guidelines to design more advanced QGLMs with computational advantages. For instance, a recent study \cite{gili2022evaluating} argued to take generalization as a measure to quantify the capability of GLMs and QGLMs in generating unseen quality and diverse data.  Nevertheless, contrary to quantum discriminative learning    \cite{abbas2020power,banchi2021generalization,du2021efficient,huang2021information,qian2021dilemma,liu2022analytic,caro2021generalization}, very few studies have attempted to unveil the learnability of QGLMs, caused by the following reasons.

The main difficulty in studying the generalization and the expressivity of QGLMs is that, unlike (quantum) discriminative models, the definition of generalization in (quantum) generative learning is vague under different models' settings.  As a result, prior literature only focuses on the specified QGLMs. Concretely, Ref.~\cite{gao2022enhancing} proved a theoretical proof of expressivity enhancement from Bayesian networks to their quantum counterpart.  Moreover, \citet{du2022theory} analyzed the generalization of QCBMs and QGANs  under the maximum mean discrepancy loss and demonstrated the potential advantages of these QGLMs. However, less is known about the generalization of QGLMs in terms of different objective functions (e.g., KL divergence) and the varied protocols (e.g., QRBMs and QAE).

The hardness in investigating the trainability of QGLMs originates from the non-convex loss landscape for the objective function. When the number of trainable parameters becomes large, the analysis for the universal convergence behavior toward the global minima is intractable. This further incurs two unsolved but crucial issues. First, how to mitigate the barren plateaus phenomenon of QGLM in the training procedure? Concisely, barren plateaus refer that the gradients of QNN could be exponentially vanished when the number of qubits and the circuit depth are linearly increased. Although \citet{kieferova2021quantum} proved that unbounded loss may enable the absence of barren plateaus for QBMs and Unitary QNNs, the strategies to alleviate barren plateaus for other QGLMs with the varied loss functions are rather unclear. Second, how to ensure the optimization can locate the optimal or near-optimal result? A large discrepancy between the estimated and the optimal results hint the large difference between the estimated and the target distribution, which may make the potential advantages of QGLMs ambiguous.

\textit{\underline{Noise}.}
The performance of QGLMs can be significantly degraded by the quantum hardware imperfection such as gate noise and measurement error. As discussed in \citet{stilck2021limitations}, when the level of noise exceeds a threshold, the power of near-term quantum computers is inferior to the classical computers. Moreover, quantum system noise may result into the diverged optimization \cite{du2021learnability}. Consequently, the unavoidable noise   forces us to redesign QGLMs, i.e., how to improve the robustness of QGLMs but keep the superiority of QGLMs? That is, when executed on NISQ machines, the output of QGLM is expected to be near to the solution under the ideal setting. Conventional approach is  adopting the hardware-efficient Ans\"atze with a shallow circuit depth to avoid the negative effects of quantum noise, while the price to pay is suppressing the power of QGLMs and diminishing the  computational advantages of QGLMs. More clever methods are demanded.

\subsection{Prospects}
There are several promising directions for the future study of quantum generative learning.

\textit{\underline{Read-in \& Read-out bottleneck}}. The query of efficient state preparation methods is ubiquitous in quantum computing, which can tackle the read-in bottleneck of QGLMs as well as other quantum algorithms. Recently, \citet{zhang2022quantum} analyzed the optimal circuit depth in quantum state preparation. Namely, given an exponential number of ancillary qubits, any $N$-qubit quantum state can be prepared with $\Theta(N)$ depth circuit using only single- and two-qubit gates. Moreover, when the state is sparse with $d$ non-zero entries, the circuit depth can be reduced to $\Theta(\log(Nd))$ with $O(Nd\log d)$ ancillary qubits.   It is intrigued to generalize this result into the NISQ regime, i.e., under the tolerable error, is any  more efficient amplitude encoding method to load the classical data into quantum states? 

As for mitigating the read-out bottleneck of QGLMs, there are two potential solutions. The first one is extending the ideas developed in the fault-tolerant regime to the NISQ regime \cite{zhang2021quantum}. Another one is integrating random measurement techniques \cite{elben2022randomized}, e.g., shadow tomography \cite{huang2020predicting} and measurement grouping \cite{o2016scalable} into QGLMs, where the target data can be extracted by the post-processing effectively.

\textit{\underline{Learning machines}.} To unveil the learnability of QGLM in terms of expressivity,   generalization, and trainability, we can follow the path of quantum discriminative learning. For instance, \citet{du2022theory} have utilized the tool in statistical learning--- covering number, to quantify the generalization of  QGANs under MMD loss. Concisely, they demonstrate how the number of training data, the architecture of QNNs, and the encoding methods effect the generalization of QGANs. Notably, such a tool has also been utilized in analyzing the expressivity and the generalization of QNNs in quantum discriminative learning \cite{du2021efficient}. It is intrigued to exploit many other advanced statistical tools to understand the capabilities and limitations of QGLMs. 

A philosophy in devising powerful underparameterized QGLMs is controlling the expressivity to be moderate. That is, the optimal solution is covered by the hypothesis space represented by QGLMs, while the size of this space is restricted to be relatively small. As such, the estimated solution can well approximate the optimal solution with both low empirical risk and low generalization error. Two leading paradigms in controlling the expressivity of QGLMs are quantum circuit architecture search \cite{du2020quantum} and injecting prior knowledge such as invariance in Ansatz design \cite{larocca2022group}. In a nutshell, the former continuously discards the unintended hypothesis functions in the training stage \cite{linghu2022quantum,grimsley2019adaptive,du2020quantum,chivilikhin2020mog,li2020quantum,ostaszewski2021structure}, while the latter discards the unintended hypothesis functions in the initialization stage \cite{larocca2022group,skolik2022equivariant,meyer2022exploiting}. In this regard, an interesting future direction is combining these two paradigms to further enhance the power of QGLMs. Meanwhile, the slimmed parameter space contributes to the alleviation of barren plateaus phenomenon.

Besides delving into underparameterized QGLMs, it is nontrival to unveil the abilities of overparameterized QGLMs. Recent studies have shown the optimal training behavior of the overparameterized quantum discriminative learning models \cite{liu2022analytic}. Nevertheless, wether overparameterized QGLMs can preserve this desired property or not deserves to be further explored. Meanwhile, whether the initialization strategies designed for underparameterized quantum discriminative learning models are suitable for overparameterized QGLMs is unclear \cite{grant2019initialization,mitarai2020quadratic,dborin2022matrix,zhang2022gaussian}.

All in all, the developed theories related to the learnability of QGLM should indicate how different Ans\"atze, loss functions, and optimizers determine the performance of QGLMs and  contribute to pursue computational advantages.

\textit{\underline{Noise}.} To suppress the imperfection of NISQ machine, two possible ways are developing error mitigation techniques for QGLMs \cite{temme2017error,li2017efficient,ofek2016extending,andersen2020repeated,endo2021hybrid} and designing problem-specific and hardware-oriented Ans\"atze \cite{du2020quantum,bilkis2021semi,tang2021qubit,amaro2022filtering}. The former aims to extract noise-free results from the noisy results, which could help compensate for the errors induced by readout and the quantum circuit itself \cite{Hamilton2020ErrormitigatedDC}. Refs.~\cite{benedetti2019generative, coyle2020quantum} indicate a rough connection between a higher entanglement entropy \cite{sim2019expressibility, meyer2002global} and a better performance of QCBM.  The latter aims to design problem-oriented and hardware-specific Ansatz to suppress the effects of noise.  The effectiveness of these two approaches has been validated in the regime of quantum discriminative learning, whereas little is known about QGLMs.  

\ifCLASSOPTIONcompsoc
    \section*{Acknowledgments}
\else
  \section*{Acknowledgment}
\fi
This work was partially supported by the National Natural Science Foundation of China (No. 91948303-1, No. 61803375, No. 12002380, No. 62106278, No. 62101575, No. 61906210) and the National University of Defense Technology Foundation (No. ZK20-52).
 
\ifCLASSOPTIONcaptionsoff
  \newpage
\fi

%\bibliographystyle{IEEEtranN}
%\bibliography{mybibliography}
% 

% Generated by IEEEtranN.bst, version: 1.14 (2015/08/26)

\end{document}